\documentclass[aps,pra,showkeys,preprintnumbers,twocolumn,nobalancelastpage,longbibliography]{revtex4-1}
\usepackage{lmodern}
\usepackage[utf8]{inputenc}
\usepackage[T1]{fontenc}
\usepackage[english]{babel}
\usepackage{url}
\usepackage{graphicx}
\usepackage{dsfont}
\usepackage{amsmath,amssymb}
\usepackage{hyperref}
\usepackage{braket}
\usepackage{enumerate}
\usepackage{float}
\usepackage{bm}
\usepackage{textcomp}

\usepackage{skull}
\usepackage{siunitx}
\usepackage{todonotes}
\usepackage{tikz}
\usetikzlibrary{quantikz}

\newcommand{\zero}{\lstick{$\ket{0}$}}

\begin{document}
\title{Towards a realistic GaAs-spin qubit device for a classical error-corrected quantum memory}
\author{Manuel Rispler}
\affiliation{QuTech, Delft University of Technology, Lorentzweg 1, 2628 CJ Delft, The Netherlands}
\affiliation{JARA Institute for Quantum Information, Forschungszentrum J\"ulich GmbH, 52428 J\"ulich, Germany}
\author{Pascal Cerfontaine}
\affiliation{JARA Institute for Quantum Information, RWTH Aachen University, 52074 Aachen, Germany}
\author{Veit Langrock}
\affiliation{JARA Institute for Quantum Information, Forschungszentrum J\"ulich GmbH, 52428 J\"ulich, Germany}
\author{Barbara M. Terhal}
\affiliation{QuTech, Delft University of Technology, Lorentzweg 1, 2628 CJ Delft, The Netherlands}
\affiliation{JARA Institute for Quantum Information, Forschungszentrum J\"ulich GmbH, 52428 J\"ulich, Germany}
\affiliation{EEMCS, Delft University of Technology, Mekelweg 4, 2628 CD Delft, The Netherlands}

\begin{abstract} Based on numerically-optimized real-device gates and parameters we study the performance of the phase-flip (repetition) code on a linear array of Gallium Arsenide (GaAs) quantum dots hosting singlet-triplet qubits. We first examine the expected performance of the code using simple error models of circuit-level and phenomenological noise, reporting, for example, a circuit-level depolarizing noise threshold of approximately $3\%$. We then perform density-matrix simulations using a maximum-likelihood and minimum-weight matching decoder to study the effect of real-device dephasing, read-out error, quasi-static as well as fast gate noise. Considering the trade-off between qubit read-out error and dephasing time ($T_2$) over measurement time, we identify a sub-threshold region for the phase-flip code which lies within experimental reach.
 \end{abstract}

\maketitle

\tableofcontents

\section{Introduction}
While the theoretical cornerstones of robust quantum computing have been established in the two previous decades in the form of a theory of quantum error correction and fault tolerance -- see e.g.~\cite{terhal:review} and references therein -- actual experiments putting this body of research to use are a matter of active research. Until now, the use of a repetition code has been demonstrated in various platforms, such as superconducting qubits \cite{Kelly2014}, and the use of error detection via the 4-qubit code has been demonstrated in ion-trap \cite{linke+:detect} and superconducting qubits \cite{takita:detect,Vuillot2017, HF:detect}. At the same time, the use of bosonic quantum error correction has led to qubits with enhanced lifetimes \cite{ofek+:cat}.

In this work, we model an experimental setup with the aim of demonstrating error suppression of Pauli-$Z$ errors in a spin qubit architecture that holds the prospect of scalability. As is well known, the current state-of-the-experimental art on spin qubits allows for 1D connectivity, while 2D connectivity and control is more challenging. Indeed, a big challenge in scaling up spin qubits is the `fan-out' problem: each quantum dot needs several control lines for defining the dot potential and operating the qubit. This requirement severely limits the capability of tightly packing many qubits onto a chip, especially in a two-dimensional fashion, see e.g.~\cite{franke+:rent} for discussion. The 2D connectivity required for the surface code is highly non-trivial \cite{crossbar}: promising realizations of long-range two-qubit gates by electron shuttling \cite{Taylor2005, Fujita2017} or mediated via superconducting resonators \cite{Harvey2018} are under active but incomplete development.

Identifying dephasing as the dominant qubit noise process leads us to suggest a one-dimensional layout on which we can operate the phase-flip repetition code.  In itself this will not give us a full logical qubit, as it will enhance the $X$-error rate of the encoded qubit while lowering the $Z$-error rate. However, it would allow a demonstration of decoding as needed for the surface code. In addition, the tolerable noise threshold error rate is more relaxed as compared to the surface code \cite{terhal:review}. This milestone -- making a 1D array which realizes the repetition code and demonstrating that the logical error decreases with increasing code distance -- has been reached for superconducting devices in \cite{Kelly2014}. While challenging, we will show that such a device would be currently experimentally feasible for singlet-triplet qubits, even in GaAs, in terms of fabrication and operation, assuming the two-qubit device performance as investigated by Ref.~\,\citenum{Cerfontaine2019a}, without the need for additional elements.  We show this by numerically running decoders on repeated error correcting cycles of the phase-flip code, subject to phenomenological and circuit-level noise, as well as realistic device-specific noise, which we simulate using full density-matrix simulations. Previously, such density-matrix simulations and fine-tuned decoders have been very informative in analyzing the performance of small surface codes, such as Surface-17 \cite{Tomita2014} in superconducting devices \cite{OBrien2017}. Even though our study is focused on GaAs qubits, a very similar analysis could be done for Si-based qubits: we review and compare some of the elementary-component specs.\ for Si-based qubits in Section \ref{sec:si}.

We note that previous work exploring the use of a 1D quantum dot array for making a logical qubit has been done in \cite{Jones2018}. However, in this previous work the focus was partially on the 4-qubit code and its concatenation, which, executed with purely 1D connectivity, has a depressingly low threshold ($\sim 10^{-4}$). Here instead we focus solely on the repetition code with natural noise (i.e. no artificially inserted errors) and examine logical error rates with growing array size.

In the next Section \ref{sec:qubits} we review the singlet-triplet qubit in GaAs and our modeling of relevant hardware components, including the proposed device in Section \ref{sec:array}. 
In Section \ref{sec:code} we review aspects of the phase-flip code and discuss decoding methods and results using simple phenomenological and circuit level error models. This sets the stage for the study and interpretation of the effects of real noise modeled using full density-matrix simulations in Section \ref{sec:num}. In Section \ref{sec:num} we study the trade-off between readout error and an effective dephasing error rate, the effect of enhanced gate-noise, the effect of stochastifying the noise via the Pauli twirl approximation, and discuss leakage and leakage reduction (not included in numerical simulations).  In Section \ref{sec:enhanc} we also numerically present and discuss the enhanced $X$-error rate. In  \ref{sec:outlook} we provide a qualitative discussion about going beyond the 1D phase-flip code: a phase-flip repetition code could be the starting point for a surface code architecture, akin to using the phase-flip code as the bottom code in \cite{Aliferis2009, AP:bias}. We will argue however that the advantage of this approach is not immediate, but depends on various current unknowns such as the noise-bias $\&$ fidelity of long- or short-range two-qubit gates.

\section{The singlet-triplet qubit in Gallium Arsenide}
\label{sec:qubits}

The idea of using the spin degree of freedom of a single electron is among the first proposed physical realizations of a qubit \cite{DL}. One suitable environment to store, control and readout single electrons with all-electric control is offered by the platform of gate-defined quantum dots in semiconductor heterostructures. The heterostructure is built such that a two-dimensional electron gas emerges, from which we can load single electrons into quantum dots that are formed by defining a confining potential through applying voltages across gates. Two prominent materials are silicon \cite{fogarty+:si} and Gallium Arsenide (GaAs). Since device fabrication in GaAs is straightforward and we have validated noise models at hand, we focus on that implementation. We comment on the (close) relation to silicon spin qubits in Section \ref{sec:si}. We will use the singlet-triplet qubit encoding, where the qubit states are chosen as the $m_z=0$ subspace of two electron spins residing in two adjacent quantum dots \cite{Petta2005}. This has several advantages: the encoding allows the qubit subspace to be less sensitive to the nuclear spin background present in the host material (which is especially prominent in GaAs, it can be mitigated in silicon by using purified $^{28} \mathrm{Si}$). The singlet-triplet encoding furthermore makes all-electrical qubit control possible by virtue of the Pauli exclusion principle, which is the workhorse behind initialization, gate operations via exchange interaction (provided a finite magnetic field gradient maintained by other means, see Section \ref{sec:operations}) and readout via Pauli spin blockade. Since these operations are also what introduces noise to the system, we will briefly discuss these operations in this section. The states $\ket{0}$ and $\ket{1}$ of this qubit are, --expressed in the usual spin up/down notation--, the states $\ket{0} \equiv \ket{\uparrow\downarrow}$ and $\ket{1} \equiv \ket{\downarrow\uparrow}$ states in the $m_z=0$ sub-space of the two-electron spin wave-function. The two remaining triplet states ($T_{+}=\ket{\uparrow\uparrow}$ and $T_{-}=\ket{\downarrow\downarrow}$) are energetically Zeeman-split off by applying a static global magnetic field. They can, to a good approximation, be neglected \cite{Hanson2007}, but play a role in qubit leakage (see Section \ref{sec:leakage}). Control over this qubit is established via the exchange interaction between the two electrons in the adjacent dots, which stems from a virtual hopping process between the two dots. This virtual hopping is only allowed for the singlet due to the Pauli principle, creating a small energy difference between the two states, which gives an effective Heisenberg type interaction. The strength of this interaction can be varied by shifting the dot potentials (i.e. voltages) with respect to another (the ``detuning''). The exchange coupling strength of the effective Hamiltonian (see Eq.~(\ref{Qubit_Hamiltonian}) can be changed rapidly on the order of nanoseconds through the use of arbitrary waveform generators. The second single-qubit control axis is given by magnetic field gradients between neighboring dots. In the case of GaAs, the nuclear background field can be locally polarized by so-called dynamic nuclear polarization (DNP) \cite{Bluhm2010}, creating the desired gradient field. Fast control of the exchange interaction allows for gate operations with fidelities above $99\%$ by numerical optimization of the pulse sequence \cite{Cerfontaine2014, Cerfontaine2019b}. We go into some more detail on gate operations in Subsection \ref{sec:operations}. Qubit relaxation is quite strongly suppressed as the qubit frequency sits at a relatively low density of states in the environment, thus leading to a high $T_1$ time, see Table \ref{tab:times}. In contrast to this, the bare coherence $T_2^*$ time can be lower than $100$ ns, but this can be remedied by echoing techniques (Hahn echo, CPMG), which lead to effective $T_2$ times in the order of \textmu s with recent experimental results going as high as $870 $\textmu s \cite{Malinowski2016}, see Table \ref{tab:times}. The language of coherence and relaxation times is common, when characterizing qubits experimentally, which is why we stick to this language. Let us however remark that ultimately the parameters relevant to error correction are the error rates (cf. Eq. \ref{eq:phasefliperrorrate}). While the two are closely related, the exact relationship depends on the underlying noise model. For $T_1$ and $T_2$ this is a Markovian model, which implicitly assumes an exponentially decaying state fidelity, thereby potentially overestimating error rates when non-Markovian low frequency noise is present (and not completely filtered out by the respective echoing technique), such that in some regime the state fidelity decays slower than exponentially \cite{Dial2013}. For a detailed discussion see e.g. \cite{Cywinski2008}. In the other extreme, very high coherence times are limited by the gate noise induced by the gates that are employed for the echoing pulses themselves. Two-qubit gates are the focus of the current research effort and recent results of accurately modeling the two-qubit operation analogous to the single qubit case suggest operability at the same high fidelities as the single qubit gates \cite{Cerfontaine2019a}. 

\begin{table}[tb!]
\begin{center}
\begin{tabular}{c|c|c}
  Time & Values (range) & Num. Simulation \\
  \hline
  $T_1$ & $\approx 2$ms \cite{Dial2013} & fixed \\ 
  $T_2^*$  & 100ns \cite{Bluhm2010} & lower limit on $T_2$ \\
  $T_2$ & 870 \textmu s \cite{Malinowski2016} & varied versus $t_{\rm readout}$ \\
  $t_\mathrm{1q-gate}$ & 20ns \cite{Cerfontaine2019a} & fixed\\
  $t_\mathrm{2q-gate}$ & 50ns \cite{Cerfontaine2019a} & fixed \\
  $t_{\mathrm{readout}}$ & 1\textmu s \cite{Nakajima2017} & fixed \\
\end{tabular}
\caption{Time duration of gates and measurement on GaAs singlet-triplet qubits. The last column indicates whether the parameter is fixed or varied in the numerical simulation.}
\label{tab:times}

\end{center}
\end{table}

\begin{table}[tb!]
\begin{center}
\begin{tabular}{c|c|c}
  Component & Infid. $1-\mathcal{F}$ & Num. Simul. \\
  \hline
  Single-qubit gate & $0.1-5.0 \hspace{0.5pt} \times 10^{-3}$ \cite{Cerfontaine2019a, Cerfontaine2019b} & varied \\
  Two-qubit gate & $0.1-5.0 \hspace{0.5pt} \times 10^{-3}$ \cite{Cerfontaine2019a} & varied \\
  Qubit measurement & $5\times 10^{-3}-1 \times 10^{-1}$ \cite{Nakajima2017, Barthel2009}  & varied \\
    Qubit initialization & $3\times 10^{-2} $  & varied \\
\end{tabular}
\caption{Error rates of GaAs singlet-triplet qubits. Here $\mathcal{F}$ is the average gate and readout fidelity respectively. The single-qubit gate numbers are based on numerical models and experimental randomized benchmarking, while the two-qubit gate numbers are based on numerical models so far. Note that we do not vary the fidelity itself but the detailed Hamiltonian model underlying the gate (see Section \ref{sec:operations}).}
\label{tab:error}
\end{center}
\end{table}

\subsection{Qubit initialization}
\label{sec:initialization}
A natural qubit initialization state is the singlet. Using the usual charge notation referring to the number of electrons in the respective dots, we start in a (1,1) charge configuration at small detuning. The dots are then largely detuned so that an electron will tunnel from one dot to the other irrespective of the spin state, yielding a (2,0) charge configuration. Next, the dot potentials are set to a configuration where a triplet state would exchange one electron with the reservoir, resulting in a (1,0) charge state, whereas a singlet state would remain in (2,0). This in effect makes sure we end up in a singlet (2,0) state, which can thus be initialized by fast electron exchange with the lead by waiting in this configuration for a time on the order of tens of nanoseconds \cite{Petta2005, Botzem2018}. For gate operations, one typically moves to smaller detuning. The gates done in \cite{Cerfontaine2019a}, which will be explained in the subsequent section, assume an incoming small detuning baseline at the start of the pulse sequence in order to deal with pulse transients caused by the finite bandwidth of the voltage pulses used for qubit control (for details see \cite{Cerfontaine2019a}). In order to reach this qubit operation point after the singlet initialization, one can either adiabatically decrease the detuning, such that the singlet hybridizes with the triplet state $T_0$ into the state $\ket{\uparrow\downarrow} = \ket{0}$ (or respectively $\ket{\downarrow\uparrow}$ depending on the direction of the magnetic field gradient). Adiabatic ramping takes on the order of 200 ns and the likely error type is accidentally initializing the higher qubit state $\ket{1}$, which we model as a bit-flip channel with varying error probability in the few percent range. An alternative way would be to diabatically move to small detuning, which could be done in about one nanosecond, preserving the singlet state $\ket{S} = \frac{1}{\sqrt{2}}\left(\ket{0} - \ket{1}\right)$. The gate simulations we use in the present work and their associated experimental results always employ(ed) adiabatic ramping (due to the pulse transients as mentioned above). In contrast to this, diabatic ramping could potentially be more appealing due to its rapidity, furthermore already preparing the ancilla in a superposition state needed for the syndrome extraction we employ for the error correcting code. Since the effect of diabatically moving to the qubit operation point lacks experimental characterization, we cover both methods in our simulations. We fix the initialization time in the circuit model (see Figs. \ref{fig:checkcircuit} and \ref{fig:circuit_t2_vs_readout}) to 20 ns, any longer initialization time can be subsumed under the measurement time, which will be a variable in the simulation results, thereby leaving the choice of initialization method open to experimental characterization and preference. Simulationwise, both initialization methods, adiabatic ramping followed by a $R_Y(\pi/2)$-rotation gate and directly initializing a singlet, are equivalent, the missing bit-flip for initializing a $\ket{+}$ instead of the resulting singlet state $\ket{-}$ from diabatic ramping can be absorbed into the interpretation of the ancilla measurement by flipping the measurement bit outcome. Likewise to the varying initialization time, for the purposes of this paper, the initialization error will be subsumed under readout error, since for all circuits used in this work, the (bit-flip) error can be propagated through the quantum circuit to the measurement location (see Fig. \ref{fig:checkcircuit}). This can be seen by the fact that the $R_Y(\pi/2)$-rotations convert between $X$ and $Z$ (modulo prefactors), so that a bit-flip error $X$ converts into a $Z$ error which commutes with the action of the CNOT gate, and then turns back into a $X$ error before reaching the measurement location.

\subsection{Gate operations}
\label{sec:operations}

Most of the properties of the singlet-triplet qubit that we consider can be grasped by an effective Hamiltonian that describes the spin dynamics of two qubits, that is, four dots, each hosting a single electron with spin vector $\bm{\sigma}^{(i)}$, $i=1,\ldots, 4$, given by
\begin{equation}
  H(\bm{\epsilon}, \bm{b}) =  \frac{1}{4}\sum_{j=1}^3J_{j,j+1}(\epsilon_{j,j+1})\bm{\sigma}^{(j)} \cdot \bm{\sigma}^{(j+1)} + \frac{1}{2}\sum_{i=1}^4B_i(\bm{b})\sigma_z^{(i)}.
  \label{Qubit_Hamiltonian}
\end{equation}
Let us explain the terms entering this Hamiltonian. The first term is the Heisenberg interaction Hamiltonian where $\epsilon_{ij} = V_{i} -V_{j}$ denotes the dot detuning between the neighboring dots $i$ and $j$. We denote all detunings together as $\bm{\epsilon} = (\epsilon_{12},\epsilon_{23},\epsilon_{34})$. As alluded to in Section \ref{sec:qubits}, this detuning leads to an effective exchange interaction between neighboring dots (hence the sum over nearest neighbors), which is described in terms of the Pauli matrices $\bm{\sigma}^{(i)}$ and the interaction strength $J_{ij}$. The second term in the Hamiltonian Eq.~\ref{Qubit_Hamiltonian} is a typical Zeeman Hamiltonian describing the coupling of the electrons to the magnetic fields $B_i$. By writing $B_i(\bm{b})$, we allude to the fact that the relevant variables are the magnetic field gradients $b_{ij} = B_{i}-B_{j}$, again using boldface to denote the three gradients between neighboring dots: $\bm{b} = (b_{12},b_{23},b_{34})$. The loss of one variable in moving from site variables (the on-site fields) to link variables (the gradients between neighbors) can be seen by noting that the sum of all fields is irrelevant for the qubit dynamics, since it is a term coupling to $\sum_i \sigma_z^i $, which is zero on any qubit state and thus drops out of the dynamics. We give the explicit change of basis in the Appendix \ref{app:Zeeman}. The dynamics of this Hamiltonian allow for universal control, i.e. arbitrary single and two-qubit gates, provided that the magnetic field gradients $\bm{b}$ are finite. Note that we only describe the dynamics of the dots involved in the respective quantum gate, which implicitly assumes negligible cross-talk to other qubits, i.e. we can execute gates in parallel on disjoint sets of qubits. Finding optimal pulse sequences is a challenging optimization problem, we refer the reader to \cite{Cerfontaine2019a}, where the average gate fidelity was used as the target function of a numerical optimization routine in order to find good composite pulse sequences for single- and two-qubit gates. The gate unitary coming out of such a pulse sequence is given by the time-evolution operator over a time duration, namely the gate execution time $t_{\rm gate}$, which is divided into $N$ intervals of length $\Delta t$. During each individual interval the couplings are kept approximately constant, such that the total time-evolution operator is of the form
\begin{equation}
  U = \prod_{m=1}^{N} \exp(-iH(\bm{\epsilon}_m, \bm{b}) \Delta t).
  \label{time_evolution_gate}
\end{equation}

As mentioned in Section \ref{sec:qubits}, the finite magnetic field gradients $\bm{b}$ necessary for universal gates are implemented by dynamic nuclear polarization (DNP) of the background nuclear field (in silicon this is implemented instead by e.g. the use of micromagnets due to the absence of the nuclear field \cite{Takeda2019}). This poses a potential problem, since during DNP the dot cannot host a qubit while being used for polarizing the nuclear field. Reaching sufficient levels of polarization takes on the order of hundreds of milliseconds. For the purpose of running the phase-flip code we imagine first using all dots to perform DNP, then initialize the qubits, execute the rounds of error correction for the phase-flip code and finally measure all the qubits (see e.g. Fig.~\ref{fig:checkcircuit}). For the small distances of the error-correcting code in question in the present work, DNP does not have to be repeated during the circuit execution, because the runtime is on the order of tens of microseconds (number of QEC cycles times measurement integration time, see Table \ref{tab:times}) and one round of DNP enables subsequent qubit operations for time-scales of at least milliseconds (as on these time-scales the nuclear field is stable) before another round of DNP is necessary. The noise characteristics of these gates are captured by the noise affecting the control parameters: the dot detuning and the magnetic field gradient (`control noise'). We denote the detuning noise by $\delta \epsilon (t)$ and the noise on the magnetic field gradient by $\delta b$, changing the Hamiltonian in Eq.~\ref{Qubit_Hamiltonian} to one which including the effective noise processes by changing
\begin{align}
  J_{ij}(\epsilon_{ij}) \rightarrow J_{ij}(\epsilon_{ij} + \delta \epsilon_{ij})\\
  b_{ij} \rightarrow b_{ij} + \delta b_{ij}.
\end{align}
The detuning noise constitute variations on the voltages, which suffer from random telegraph noise, which is explained by charge traps in the vicinity of the quantum dot, that are periodically loaded and unloaded, leading to small spikes in the voltage signal. While still under debate, this model agrees well with experimental observation and yields $1/f$-type noise (in reference to the power spectrum) \cite{Dial2013}. For this detuning noise we use a model that encompasses fast and slow components compared to the time-scale of the execution of a gate. The magnetic field noise $\delta b $ is dominated by low-frequency noise, the underlying mechanism being nuclear spin diffusion, which is well described by slowly varying (quasi-static) noise \cite{Reilly2008,Bluhm2010}.

The effect of control noise on the level of gate execution is that we are implementing a Hamiltonian whose parameters are slightly offset, in turn leading to the actual unitary gate applied to the qubit being slightly different from the intended gate. If these control parameters are changing fast compared to gate operation times, then such noise can be well described by its quantum channel which is averaged over many noise realizations. In contrast to this, noise parameters that change slowly with respect to gate operation times lead to the computation being subject to the same systematic error over its entire time. In terms of simulating these operations, the fast noise channel can be simulated once and then acts as the same channel at any location. To model slow noise, we will instead draw noise parameters per circuit run instance and only do the averaging after sampling from this distribution on the level of the entire circuit simulation. The details of this model are given in Appendix \ref{app:noise-model-detail}. Another error channel besides control noise is leakage, which we briefly discuss here. When looking at the joint Hilbert-space of two neighboring qubits, we can see that the $m_z=0$ subspace is 6-dimensional, since it contains the two additional states $\ket{T_+ T_-}$ and $\ket{T_- T_+}$. These can be reached from a computational two-qubit state by exchanging the spin between the two dots in the middle (e.g. $\ket{\uparrow \downarrow \uparrow\downarrow} \rightarrow \ket{\uparrow \uparrow \downarrow\downarrow} $), which is not protected by the Zeeman field. However, these transitions can be avoided by making such a spin-flip energetically costly by imposing that at all times $b_{23}=B_3-B_2\gg J_{23}$ (cf. Eq.~\ref{Qubit_Hamiltonian}). Detailed two-qubit gate simulations \cite{Cerfontaine2019a} predict very low leakage rates ($\approx 10^{-4}$), which is why we focus on control noise errors in the present work, deferring leakage to Section \ref{sec:leakage}, where we explain how leakage can be incorporated into error correction via leakage reduction.\\
\subsection{Readout}

A feature of the singlet-triplet qubit is that it grants access to a high fidelity readout mechanism in the form of spin-to-charge conversion. Here one exploits the Pauli exclusion principle, which forbids the two electrons in the triplet state to be in the same dot. By going to a high detuning, it becomes energetically favorable for the singlet to have both electrons in the same dot. This way we can discriminate the qubit states by measuring the number of electrons in either of the dots, which can be done using a single-electron transistor (SET). Readout fidelity is a bottleneck in current spin qubit implementations, both in terms of the measurement integration time, i.e. the time one needs to discriminate the signal on the SET as well as in terms of the fidelity of this measurement. While current readout numbers can be as bad as milliseconds and $80\%$ fidelity \cite{Watson2018}, we will see that even these numbers are not necessarily fatal for our purposes, since the code at hand can in principle tolerate a high readout error in a certain parameter regime. On the other hand, there have been recent proposals pushing the envelope of spin qubit readout to the $99\%$ fidelity regime and integration times on the order of microseconds by using a latched readout \cite{Nakajima2017}. A dispersive qubit read-out using an on-chip resonator, taking $6$ \textmu s with fidelity $98\%$ has been achieved for Si qubits \cite{zheng+:resonator}. In Table \ref{tab:times} and Table \ref{tab:error} we summarize some of these numbers for the GaAs singlet-triplet qubit. In Table \ref{tab:si} and Section \ref{sec:si} we discuss similar numbers for Si qubits.
\subsection{Qubit dephasing and noise tradeoff}
So far, we have described gate errors and readout errors. In a typical quantum error correction protocol (including the one employed in this work) \cite{terhal:review}, to protect the information encoded in the data qubits, one extracts error information by using ancilla qubits, which are subsequently measured. During the measurement integration time, the data qubits of the error-correcting code have to wait before we can continue with the next cycle. As the measurement time is $O(100)$ times slower than the individual gate times in Table \ref{tab:times}, dephasing of the data qubits during measurement by far dominates dephasing-induced $Z$ or $X$-errors during the rest of the QEC cycle.

This slow measurement bottleneck can in principle be circumvented by supplying fresh ancillas for new QEC cycles while the current ancillas are still being measured, see e.g. the analysis in \cite{DA:slow}, or by avoiding measurement all together \cite{ercan+:meas-free}.
However, we choose to treat qubits as a scarce resource, which seems more reasonable in the near-term, and thus re-use the measurement qubits to keep the number of qubits minimal. As described above, the dominant qubit noise component is dephasing, which means that during the idling, the data qubits suffer a phase-flip error rate \cite{Sarvepalli2009}
\begin{equation}
  p(Z) = \frac{1-\exp(-t_{\rm readout} /T_2)}{2},
  \label{eq:phasefliperrorrate}
\end{equation}
where $t_{\rm readout}$ is the read-out time and $T_2$ is the dephasing time which can vary,-- depending on dynamical decoupling pulses--, from $T_2^*$ to the best values reported with dynamical decoupling during the readout.

All-in-all, the efficacy of phase-flip error correction will depend on (1) the intensity of $Z$ errors due to the CNOT gates, (2) the ancilla qubit measurement error rate, and (3) the ratio $t_{\rm readout}/T_2$. Naturally, the latter quantity can be experimentally reduced by either shortening the read-out time or lengthening the dephasing time by more intensive dynamical decoupling (although there is a limit on how much dynamical decoupling can help as the single-qubit gates themselves introduce additional $Z$ errors).  We identify the tradeoff between the readout quality (2) and qubit dephasing (3) as the main tradeoff of the proposed experiment.
\subsection{Proposed linear dot array}

\begin{figure*}[!tb]
\includegraphics[width=0.8\textwidth]
{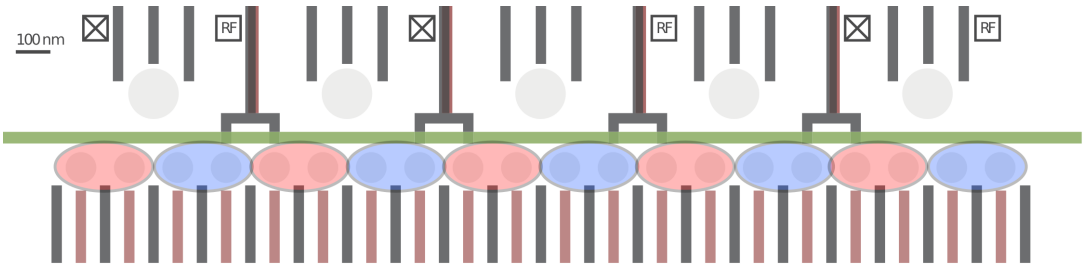}
\caption{Top-view layout of a device hosting a phase-flip code in a one-dimensional array of quantum dots. The qubits are in the center of the device, indicated by alternating red (data qubits) and blue (ancilla qubits) ellipsoids each encompassing two quantum dots, since we are using singlet-triplet qubits. The gate indicated by the central horizontal green line isolates the quantum dots from the electron reservoirs situated above. The pitchfork gates can be used to couple each ancilla and qubit to an electron reservoir, which is essential for fast initialization and readout. For the latter, the SETs (single-electron transistors) situated between the pitchfork gates can be used to detect the change of electron number in the dot. The lines from below are the gates used for defining and operating the quantum dots (cf.~\cite{Volk2019}). Note that the rightmost ancilla qubit is not needed for the phase-flip code but useful for leakage reduction (see \ref{sec:leakage}.)}
\label{fig:chip-layout}
\end{figure*}
\label{sec:array}
To operate the phase-flip code, we imagine a one-dimensional array of alternating data and ancilla qubits, which we realize by a linear array of quantum dots as shown in Fig.~\ref{fig:chip-layout}. This layout is modified with respect to previous linear designs used in several groups \cite{Volk2019, Ito2016, Zajac2016} to allow singlet-triplet qubit operation with dynamic nuclear polarization (DNP) \cite{Bluhm2010}.

To this end, our proposal first simplifies the fan-out of the large number of electrostatic surface gates by moving to the third dimension and stacking three gate layers vertically, isolated from each other by dielectric layers in between. This enlarges the space between adjacent gates and thus lowers the demands on the fabrication process. In Fig.~\ref{fig:chip-layout}, the lowest layer is indicated in red, the middle layer in dark gray and the top layer in green.

Similar to previous designs, the dots are defined by a long horizontal gate (on the topmost layer and thus shown in green) and so-called barrier and plunger gates coming in from the bottom (dark gray and red, respectively). The barrier gates predominantly control the tunnel barriers to adjacent dots, and to the electron reservoirs for the left-most and right-most dots. The plunger gates mainly control the dot potentials. By using bias-tees, the DC potential of these gates can be set using stable voltage sources while fast control with roughly \SI{300}{MHz} bandwidth (for manipulation, initialization and readout) can be coupled in simultaneously using arbitrary waveform generators.

In this array, qubits are encoded in two separate but tunnel-coupled quantum dots (using the singlet-triplet encoding described in Section \ref{sec:qubits}). Experiments \cite{Volk2019} have shown that such an array can reach sufficiently high tunnel couplings \cite{Botzem2018} on the order of tens of \SI{}{\micro eV}. In order to sense the charge state of each dot in the linear chain, sensing dots (SETs) are defined by gates coming in from the top of the diagram. Each SET is confined by the green horizontal gate and three additional gates. The SETs are used for readout of the two qubits next to them. They are connected to electron reservoirs via Ohmic contacts (square boxes) which are in turn connected to a ground potential (black crosses) or a RF readout circuit \cite{Reilly2007}.

In order to allow straightforward singlet-triplet qubit operation, we add the black "pitchfork" gates coming in from the top. These gates can be used to couple each qubit to an electron reservoir by metallic screening of the confinement potential of the green gate. This coupling can be used for fast initialization in a singlet state, which is essential for performing DNP. The black pitchfork gates are screened by lower-lying gates indicated in red so that the electron reservoir surrounding the SETs is not affected by their presence. This allows the potential of the pitchfork gates to be changed without a detrimental effect on the SET potentials.

Direct initialization of each qubit is useful for initializing a well defined qubit state, and for performing DNP to control magnetic field gradients between adjacent dots (see Section \ref{sec:operations}). Sufficiently high magnetic field gradients can suppress leakage to non-computational $T_-$ and $T_+$ states during two-qubit gates (see Section \ref{sec:operations} and Ref.~\cite{Cerfontaine2019a}).

\subsection{Silicon-based qubits}
\label{sec:si}

While we picked Gallium Arsenide for this particular realization, our results are for the most part also informative for silicon qubits, with the Silicon- $ST_0$-qubit being the closest cousin to our setup. The gate simulations that were done in \cite{Cerfontaine2019a} also discuss this relationship and conclude that similar gate fidelities can be expected in silicon. We list some typical parameters for silicon qubits in the Table \ref{tab:si} for comparison (note that there are different types of silicon qubits, see references). It is apparent that the noise bias towards decoherence $T_2$ versus relaxation $T_1$ is equally prominent in silicon. The main reason for using (purified) silicon is the absence of the nuclear field, which allows for substantially higher relaxation and decoherence times. However, this has the drawback of losing the possibility for dynamic nuclear polarization, which is typically replaced by mounting micromagnets on the sample, which still poses questions to scalability. Gate times are substantially longer, to some extent remedied by advantages in relaxation and decoherence times. An advantage towards dealing with leakage is that single electron spins can be used as qubits; this has however the major drawback that the readout is not high-fidelity. This suggests that a hybrid device could be used with ancilla qubits being singlet-triplet qubits and data qubits embodied by single spins.
 
\begin{table}[htb]
\begin{center}
\begin{tabular}{c|c|c}
  Component & Values (range) & type \\
  \hline
Gate times & $\approx 1$\textmu s \cite{HuangDzurak2019} & LD \\ 
  2-qu. gate $1-\mathcal{F}$ (Exp.)& $2\times10^{-2}$ \cite{HuangDzurak2019} & LD\\
  2-qu. gate $1-\mathcal{F}$ (Th.) & $1\times10^{-4}$ \cite{Cerfontaine2019a} & $ST_0$\\
  $T_2$ & 1ms \cite{Yoneda2017} & LD\\
  $T_1$ & >1s \cite{Hollmann2019} & LD\\
  Measurement time & 1-6 \textmu s \cite{Schaal2019,zheng+:resonator} & $ST_0$\\
  Meas. error rate & $2\times10^{-2}-3\times10^{-3}$ \cite{Schaal2019,zheng+:resonator} & $ST_0$\\
 Meas. error rate &  $2\times10^{-1}$ \cite{Watson2018} & LD
\end{tabular}
\end{center}
\caption{Comparative table of average gate fidelities and operation times for silicon qubits found in the literature. Since there are several qubit types, the type is indicated as LD (Loss-DiVincenzo) for the bare single electron spin qubit and as ($ST_0$) for the singlet-triplet qubit. Note that the two-qubit gate infidelity for LD in silicon is the experimental number so far and can be expected to decrease, at least on the basis of theoretical models.}
\label{tab:si}
\end{table}
\section{Phase-flip code}
\label{sec:code}

The phase-flip code is simply the classical repetition code with $n$ qubits: instead of correcting bit-flips we protect against phase-flips by rotating the stabilizer checks. While being rather simplistic, this code is a small example of a topological code since it is a one-dimensional variant of the surface code. The phase-flip code can be viewed as the natural stepping stone or testbed towards implementing the surface code, as its manner of decoding is similar, see also Section \ref{sec:outlook}.

\subsection{Parity checks, logical operators, preparation and measurement}
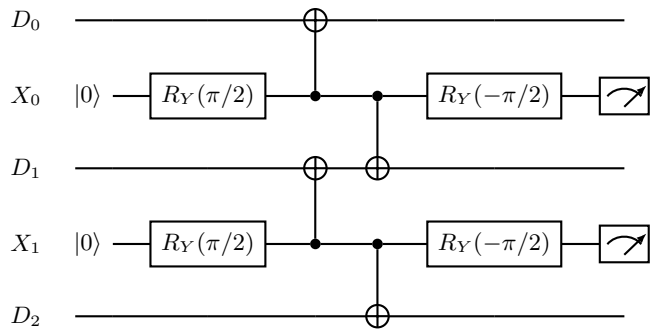
\begin{figure}[H]
\begin{tikzcd}  
  \lstick{$D_0$}\quad & \qw                &  \qw        & \targ{}   & \qw      & \qw        & \qw      \\
  \lstick{$X_0$}\quad & \lstick{$\ket{0}$} &  \gate{R_Y(\pi/2)} & \ctrl{-1} & \ctrl{1} & \gate{R_Y(-\pi/2)} & \meter{} \\
  \lstick{$D_1$}\quad & \qw                &  \qw        & \targ{}   & \targ{}  & \qw        & \qw      \\
  \lstick{$X_1$}\quad & \lstick{$\ket{0}$} &  \gate{R_Y(\pi/2)} & \ctrl{-1} & \ctrl{1} & \gate{R_Y(-\pi/2)} & \meter{} \\
  \lstick{$D_2$}\quad & \qw                &  \qw        & \qw       & \targ{}  & \qw        & \qw      \\
\end{tikzcd}
\caption{\label{fig:checkcircuit} Circuit diagram for one error correction cycle of the distance-3 phase flip code. The XX-parity checks are implemented by coupling to ancillas, which are subsequently measured to give the syndrome information.}
\end{figure}

The parity checks on $n$ qubits (with $n$ odd) on a line are the nearest-neighbor checks $S_i = X_i\otimes X_{i+1}$ for $i=0, \ldots, n-2$. The logical operators are $\overline{Z} = \bigotimes_{i=0}^{n-1} Z_i$ and $\overline{X}=X_i$ for any $i$.  These act on the logical qubit states defined as $\ket{\overline{0}} \equiv \frac{1}{\sqrt{2}}(\ket{++\ldots +}+\ket{--\ldots -})$, $\ket{\overline{1}} \equiv\frac{1}{\sqrt{2}}( \ket{++\ldots}-\ket{--\ldots -})$, 
and $\ket{\overline{+}}=\ket{++\ldots +}$, $\ket{\overline{-}}=\ket{--\ldots -}$. The code has distance $d=n$ with respect to $Z$-errors and distance $1$ with respect to $X$-errors. The parity check measurements are implemented by coupling both qubits to an ancilla through a CNOT gate, which can be done in two time-steps, in parallel on all even/odd qubits. The circuit diagram is given in Fig. \ref{fig:checkcircuit}, Here, the ancilla preparation and measurement in the $\pm$ basis is realized by using 
rotations $R_Y(\pm \pi/2)$.

A destructive logical $\overline{Z}$-measurement for the phase-flip code corresponds to measuring all qubits in the $Z$-basis and taking the product of all individual $Z_i$ outcomes. Note that each individual $Z_i$ outcome will be random, but the product of the expectation value is 1 (resp. -1) when applied to $\ket{\overline{0}}$  (resp. $\ket{\overline{1}}$). This measurement is very sensitive to measurement error as its outcome depends on all the individual outcomes. 
A destructive logical $\overline{X}$-measurement is the measurement of all qubits in the $X$-basis and taking the majority of the answers, and hence is very robust to error. The preparation of $\ket{\overline{+}}$ (resp. $\ket{\overline{-}}$) is simple as it is the preparation of each qubit in the $\ket{+}$ state (resp. $\ket{-}$).
The preparation of $\ket{\overline{0}}$ can be done by starting all qubits in the $\ket{00\ldots 0}$ state, (also) an eigenstate of $\overline{Z}$, and measuring the $S_i$ checks. These $S_i$ checks will have random outcomes and thus require single-qubit $Z$-corrections (to set their eigenvalues to $+1$) which does not affect the eigenvalue of $\overline{Z}$. In our simulations we start the data qubits in the circuit in Fig.~\ref{fig:checkcircuit} in the state $\ket{++\ldots +}$, and, after repeated rounds of quantum error correction cycles, we measure each qubit in the $X$-basis. The circuit for code size (odd) $n$ is an alternating array of $n$ data qubits with $n-1$ ancilla qubits and otherwise identical QEC cycle.

\subsection{Decoding}
\label{sec:decoding}

The decoding problem is an inference problem in which we use error information, obtained by stabilizer measurements, to infer whether a logical error has happened on the encoded qubit. Let us first make a few very general comments on decoding which apply well beyond the phase-flip code studied in this paper. 

We can describe the decoding problem as a two-player set-up in which one person is the experimenter preparing an encoded input state, performing the QEC cycles and finally deciding whether to measure the encoded qubit in the logical $\overline{X}$ or $\overline{Z}$ basis. The other player is `the decoder' who does not know the input state, nor what final measurement is chosen, but only gets all parity check measurement data ${\bf M}$ acquired during the QEC cycles. The goal of the decoder is to give the experimenter two bits $b_z=0,1$ and $b_x=0,1$ based on knowing ${\bf M}$. The experimenter uses these bits as follows. If she measures $\overline{Z}$ and $b_z=0$, she accepts the outcome of $\overline{Z}$ as the true outcome, if $b_z=1$ she flips the outcome. Similarly, if she does the $\overline{X}$ measurement, she uses the bit $b_x$ to flip the outcome or not. A slightly more general 
formulation is one in which the decoder also gets all the last destructive measurement data $M_{\rm final}$, and tells the experimenter how to read this final logical measurement using $M_{\rm final}$ and the record ${\bf M}$. The reason to think about the decoder and the experimenter as different identities is that the decoder should not be able to enhance her performance by knowing what state the experimenter started with, or by knowing in advance what measurement the experimenter will perform. In addition, she may be asked to perform the computational task of decoding {\em on-line} meaning that that there should be no time-delay due to her computation lagging behind when the experimenter finishes doing all measurements and wants to know the answer to the final logical measurement. 

The most powerful decoder could simulate the entire noisy quantum computation using a full density-matrix simulation given the best possible noise model. Naturally, this form of decoding is not scalable and defeats the purpose of quantum computing. In addition, in the two-player set-up, the computational power of such decoder has to go beyond that of quantum computing, as she has to post-select her simulation based on the measurement data that she obtains from the experimenter. We will refer to this form of decoding as maximum-likelihood decoding, as it is the best possible way of decoding. 

One can consider the performance of such a maximum-likelihood decoder for a code family of growing distance $n=d$, such as the phase-flip code against $Z$-errors. To compare the decoder's performance for different $d$, we let the experimenter execute $d$ QEC cycles before executing a final logical measurement \footnote{The reason not to examine the decoder's performance where the number of phase-flip cycles is fixed to a constant $c$, while the number of qubits $n$ in the code and its distances increases, is that one can show that in that case, the decoder will certainly be performing worse for increasing $n$. The reason is that she is more likely to interpret a connected string of qubit errors of weight $> 2c$ as two separate sequences of measurement errors which terminate in any unknown future measurement record. In other words, she will not infer any corrections for such errors whereas these errors will become prominent with increasing $n$. If the decoder performs worse with increasing $n=d$, it implies that there is no gain in using codes with larger distance $d$.}.

How do we assess the performance of the decoder? Assessing its performance first of all assumes that the noise model that we employ is an accurate description of the physical set-up, that is, we have a fairly accurate description of the full density-matrix evolution, and we can thus compare the decoder's decision with the maximum-likelihood decision.

In our case we are firstly interested in the occurrence of a logical $\overline{Z}$ error as the aim of the code is to reduce this, hence the experimenter will finally measure $\overline{X}$, i.e. measure all qubits in the $\pm$ basis. In addition, in assessing the performance of the decoder, we will imagine the last measurement step of $\overline{X}$ by the experimenter to be error-free, simply so that it does not count towards the logical error rate which is a function of the number of cycles $n$. 

Let us first give the success probability for maximum-likelihood decoding. In principle, the experimenter starts the qubits in some arbitrary encoded unknown state $\sigma$. After $n$ rounds of parity check measurements, let the output state of the data qubits be $\rho_{\bf M}^{\sigma}$ where $\bf M$ is a multi-index label for the parity-check outcomes, i.e. $\bf M $ has $\#$(stabilizers) $\times$ $\#$(QEC-cycles) entries. Now the simplest maximum-likelihood decoder assumes that the input is, say $\ket{\overline{+}}$ (instead of $\sigma$) and runs a density-matrix simulation and outputs 'flip', i.e. $b_x=1$, when in her simulation $\mathbb{P}(\overline{X}=-1|{\bf M}, \overline{+}) > \mathbb{P}(\overline{X}=1|{\bf M}, \overline{+})$. The failure probability, {\em if} the state of the experimenter was indeed $\ket{\overline{+}}$, is then
\begin{align}
\overline{P}_{\rm MLD}=\sum_{{\bf M}}\mathbb{P}({\bf M}) \;\times \notag \\ \min(\mathbb{P}(\overline{X}=-1|{\bf M}, \overline{+}),\mathbb{P}(\overline{X}=1|{\bf M}, \overline{+})).
\label{eq:mld}
\end{align}
We will take this probability as a proxy for the MLD logical failure probability for $\overline{Z}$ for arbitrary input states $\sigma$ for simplicity, as we do not expect that starting with $\ket{\overline{-}}$ would give the decoder a very different decision, or that arbitrary inputs will fare very differently. This is exactly correct when the noise model is that of depolarizing errors which act in a completely state-independent way, but it is an assumption when we implement more general noise in Section \ref{sec:num}. 
Similarly, if we would evaluate MLD decoding for a $\overline{X}$ error, we would prepare a $\ket{\overline{0}}$ state and the decoder decides $b_z=1$ when $\mathbb{P}(\overline{Z}=-1|{\bf M}, \overline{0}) > \mathbb{P}(\overline{Z}=1|{\bf M}, \overline{0})$ with corresponding logical failure probability.

\begin{figure}[tb!]
\center{\includegraphics[width=1\columnwidth]
{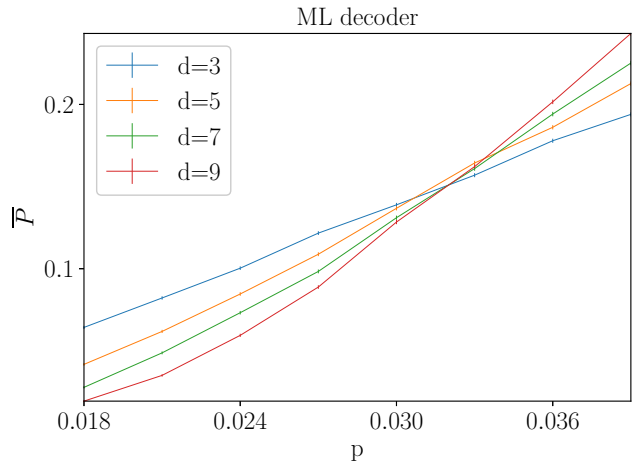}}
\caption{\label{depolarizing_logical_error_rate} Logical vs physical error rate of the phase-flip code for circuit-level depolarizing noise, using maximum likelihood decoding and plotting for increasing code distance $d$. The threshold is estimated as $p_c=3.3\%$, below which the logical $\overline{Z}$-error rate $\overline{P}$ decreases with increasing code distance. For every $d$, we run $d$ QEC cycles $N$ times with $N=10^4$.}
\label{fig:threshold-circuit}
\end{figure}

To test maximum-likelihood decoding we apply it to a standard noise model of circuit-level depolarizing noise. This model is as follows. For a single-qubit, the (symmetric) depolarizing channel is the Pauli channel
\begin{equation}
\mathcal{E}(\rho) = (1-p)\rho + \frac{p}{3} \sum_i P_i\rho P_i \label{phaseflipchannel}.
\end{equation}
Circuit-level noise means that we apply this noise channel after every single-qubit element (initialization, single-qubit gate, readout) in the circuits such as Fig.~\ref{fig:checkcircuit}. For a two-qubit gate, we apply the two-qubit version of the symmetric depolarizing channel which leaves the state alone with probability $1-p$ and applies any one of 15 two-qubit Paulis with probability $p/15$. We use full density-matrix simulations using Quantumsim software available at \cite{quantumsim} to simulate the noisy circuits, although for the noise models in this section simpler stabilizer simulations are possible. By evaluating Eq.~(\ref{eq:mld}) for different code distances and error probability $p$, we find a threshold value of the depolarizing circuit noise of $p_c=0.033$ in Fig.~\ref{fig:threshold-circuit} using maximum likelihood decoding. The threshold here is taken to be the point where the curves for different $d$ cross, which captures the asymptotic performance of the code.

To put these results in perspective and discuss the computationally-efficient method of minimum weight matching decoding, let us take a step back and first consider some simpler phenomenological noise models.
The simplest one is to assume that all parity check measurements are perfect and in each QEC round, every qubit is hit by a phase-flip error channel
\begin{equation}
\mathcal{E}(\rho) = (1-p)\rho + p Z\rho Z. \label{phaseflipchannel}
\end{equation}
After a single QEC round one can mark which stabilizers are flipped as defects and match these defects to each other or to the outside boundary, choosing the matching which minimizes the total distance between the matched defects. The matching produces a logical error when the number of phase-flip errors is at least $(d+1)/2$ (i.e. the majority is in error and - unbeknownst to the decoder - choosing the complementary matching would have been correct). Thus the logical failure probability is
\begin{equation}
  \overline{P} = \sum_{k=(d+1)/2}^{d} {d \choose k} p^k (1-p)^{d-k},
\end{equation}
which for $d\rightarrow \infty$ tends towards zero for any $p<0.5$ giving the well-known threshold value $p=0.5$ for the phase-flip code with perfect syndrome measurements.

A next model is that of phenomenological noise. In this noise model, each qubit undergoes a $Z$-flip with error probability $p$ in each QEC round and each parity check is perfect except for the outcome of the ancilla measurement being flipped with probability $q$. In this noise model we have the additional effect that defects do not necessarily correspond to actual data qubit errors anymore but can instead stem from an inaccurate measurement itself. To handle this, one needs to process syndrome measurement outcomes from one round to the next. From the record ${\bf M}$ the decoder creates a syndrome {\em defect} record by placing a defect between cycle $t$ and $t+1$ when the stabilizer measurement outcome changes from cycle $t$ to $t+1$. A measurement error at cycle $t$ will thus lead to a pair of defects at time $(t-1,t)$ and $(t,t+1)$ and pairing these defects in decoding means that we interpret it as such.  An incoming qubit phase-flip error in round $t+1$ will lead to two neighboring defects at time $t+1$ in the bulk of the lattice, which can be matched. On the boundary there is only one defect (which could be matched to the boundary). A minimum-weight matching decoder now takes this defect record and matches all defects with the goal of minimizing the total distance between the matched defects, where the distance is taken as a function of the error probability $p$ and $q$. The algorithm of minimum-weight perfect matching (MWPM) is the efficient Edmonds' Blossom algorithm \cite{Edmonds1965}.

\begin{figure}[tb!]
\center{\includegraphics[width=1\columnwidth]
{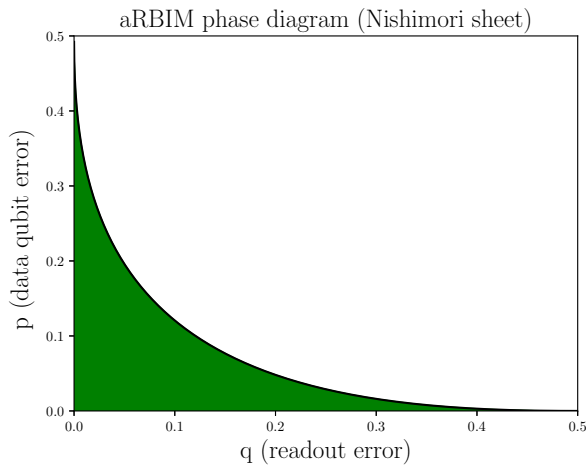}}
\caption{\label{anisotropic_RBIM_phasediagram} Phase diagram of the anisotropic Random Bond Ising model according to the Takeda-Nishimori conjecture \cite{Takeda2005} given by Eq.~\ref{Takeda-Nishimori}. The green region marks the ferromagnetic phase, which corresponds to the correctable region of the phase-flip code for anisotropic phenomenological noise.}
\label{fig:memory-phase}
\end{figure}

If we choose the measurement error rate $q$ equal to the $Z$ error rate $p$, then, except for the marginal details on the boundary, the decoding problem is known to be equivalent to the decoding problem of the toric code with perfect measurements \cite{Dennis2001}. The equivalence can be understood by seeing that the time-direction of the repetition code plays the role of the second spatial direction of the toric code. This toric code with perfect measurements has the well-known threshold value of $p= 0.11$ \cite{Wang2003} under maximum likelihood decoding. The minimum weight matching decoder performs close to optimal with a threshold of $p= 0.105$ \cite{Kawashima1999,Dennis2001}. 

The case $p \neq q$ and the optimally achievable max. likelihood threshold are relevant in understanding the numerical data for full-density matrix simulations in Section \ref{sec:num} as it effectively features the same trade-off between measurement error (error rate $q$) versus incoming error (error rate $p$). Takeda et al. ~\cite{Takeda2005} have conjectured the phase boundary separating the below-threshold to above-threshold region to lie at the following line
\begin{equation}
  H(p) + H(q) = 1,
  \label{Takeda-Nishimori}
\end{equation}
where $H(x)$ denotes the Shannon entropy $H(x)=-x\log_2(x) - (1-x)\log_2(1-x)$, see Fig.~\ref{fig:memory-phase}. When $p=q$, the condition $H(p_c)=1/2$ implies $p_c \approx 11\%$.
The threshold boundary shape in Fig.~\ref{fig:memory-phase} provides a guidance for understanding the results in the next section, e.g. Fig.~\ref{phasediagram_T2_vs_readout}. Even though the noise is more involved as it includes some level of gate noise, the main features of a trade-off between read-out error versus phase-flip error are indeed present. In Appendix \ref{sec:aRBIM} we review the underlying theory behind maximum likelihood decoding for the repetition code.

\begin{figure}[tb!]
\center{\includegraphics[width=1\columnwidth]
{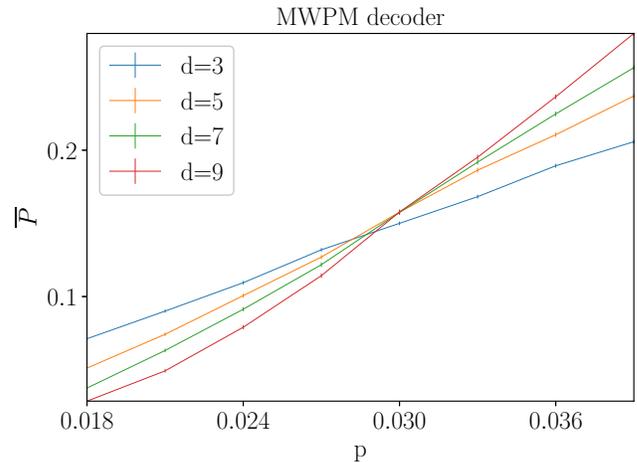}}
\caption{\label{depolarizing_logical_error_rate_MWPM} Logical vs physical error rate of the phase-flip code for the circuit-level depolarizing noise channel for increasing code distance $d$ and minimum weight decoding. The threshold with this decoder is slightly below MLD at circa $p_c=3\%$.}
\end{figure}

Let us now present our results on applying MWPM on the {\em circuit-level depolarizing} noise model to see the discrepancy with ML decoding. In order to decide how to best match up the syndrome defects, we assign equal weight to both time-like (ancilla) and space-like (data) errors, i.e. for decoding we essentially assume a phenomenological noise model with equal measurement as incoming error rate $p=q$. We then perform the matching of a given syndrome graph with the weights given by this noise model by using a standard implementation of the Blossom algorithm in the Python package networkx \cite{networkx}. As shown in Fig. \ref{depolarizing_logical_error_rate_MWPM}, the logical error rates cross at a value of $p=3\%$, which shows that MWPM with a rather simple noise model performs close to optimal for circuit level depolarizing noise. This good performance might be not too surprising given the low (=2) weight parity checks of the phase-flip code. The inclusion of so-called vertical hooks \cite{Dennis2001,Kelly2014} for the matching algorithm might bring the MWPM threshold closer to the optimal MLD threshold, however we chose not to investigate this slight difference further. For the surface code, the question of the timing of errors (i.e. at which point in the parity check circuit they occur) is much more sensitive due to the higher weight parity checks and the fact that $X$- and $Z$-checks are acting on the same qubits, which can lead to problematic error patterns (notably single ancilla errors spreading to several data qubit errors) \cite{Dennis2001}. In a phase-flip code the CNOT gates cannot propagate ancilla errors to data qubit errors due to the structure of the parity checks.

\section{Device-specific full density-matrix simulations}
\label{sec:num}

\subsection{Full density-matrix circuit simulation}

\begin{figure*}[htb]
\newcommand{\NoiseTwo}{\gate[wires=2,style={fill=red!20}]{\mathcal{N}}}
\newcommand{\NoiseOne}{\gate[style={fill=red!20}]{\mathcal{N}}}
\newcommand{\NoiseOneBlue}{\gate[style={fill=blue!10}]{\mathcal{N}}}
\newcommand{\Ttwo}{\gate[style={fill=blue!10}]{T_2}[4cm]}
\center{\begin{tikzcd}  
 \lstick{$D_0$}\quad &\qw     &\qw                        &\qw       &\targ{}   &\NoiseTwo &\qw       &\qw       &\qw                        &\qw       &\ \ldots \ \qw &\Ttwo &\qw                           & \qw \\
 \lstick{$X_0$}\quad &\zero   & \gate{R_Y(\frac{\pi}{2})} &\NoiseOne &\ctrl{-1} &          &\ctrl{1}  &\NoiseTwo &\gate{R_Y(-\frac{\pi}{2})} &\NoiseOne &\ \ldots \ \qw &\NoiseOneBlue &\meter[style={fill=blue!10}]{} &    \\
 \lstick{$D_1$}\quad &\qw     &\qw                        &\qw       &\targ{}   &\NoiseTwo &\targ{}   &          &\qw                        &\qw       &\ \ldots \ \qw &\Ttwo &\qw                            &\qw \\
 \lstick{$X_1$}\quad &\zero   & \gate{R_Y(\frac{\pi}{2})} &\NoiseOne &\ctrl{-1} &          &\ctrl{1}  &\NoiseTwo &\gate{R_Y(-\frac{\pi}{2})} &\NoiseOne &\ \ldots \ \qw &\NoiseOneBlue &\meter[style={fill=blue!10}]{} &    \\
 \lstick{$D_2$}\quad &\qw     &\qw                        &\qw       &\qw       &\qw       &\targ{}   &          &\qw                        &\qw       &\ \ldots \ \qw  &\Ttwo &\qw                            &\qw \\
\end{tikzcd}
}
\caption{\label{fig:circuit_t2_vs_readout} Circuit diagram for one noisy cycle for the distance-3 phase flip code. We use red and blue to discriminate between errors happening during parity check gate executions and errors happening during the measurement of the ancilla qubits. The red boxes $\mathcal{N}$ denote the gate noise channels on the single and two-qubit gates and the initialization error, blue boxes indicate the noise processes during readout. Time is not to scale, indicated by dotted qubit wires: the measurement takes much longer than the gate execution, during this whole time dephasing happens on the data qubits indicated by the blue gate $T_2$. The measurement is faulty, modeled by a bit-flip channel preceding the measurement. Our simulation contains the effect of $T_2$ over the full time between gate executions of consecutive QEC cycles (readout-time + gate-times), however the readout time is dominant, since all data qubits in the bulk otherwise only idle during initialization (20 ns, see Section \ref{sec:initialization}) and the $Y$-rotations of the ancillas ($2\times 20$ ns). The two boundary qubits idle an additional $50$ ns during the CNOT gates. Qubit relaxation is not indicated for readability since it is negligible but it is included in the simulation.}
\end{figure*}
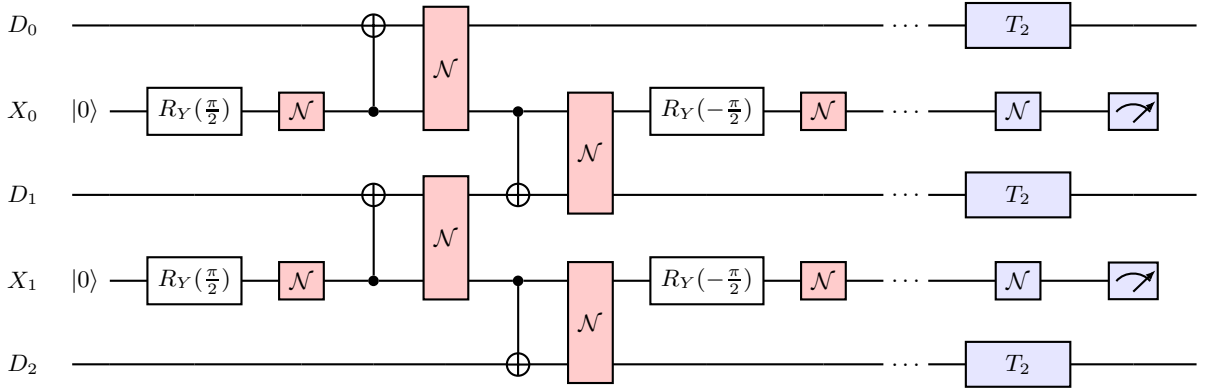

Using the Quantumsim framework \cite{quantumsim} we build a numerical simulation of the quantum circuit outlined in Fig. \ref{fig:circuit_t2_vs_readout}, where all the components are faulty, i.e. the gates are given by realistic noisy channels, the qubits decohere and the initialization and measurement can be imperfect, according to Tables \ref{tab:times} and \ref{tab:error}. In particular, the gates are derived from the qubit Hamiltonian $H(\bm{\epsilon}+\delta \bm{\epsilon}, {\bf b}+\delta {\bf b})$ with the magnetic field gradients ${\bf b}$ and the detunings $\bm{\epsilon}$ between the dots (Eq. \ref{Qubit_Hamiltonian}) to which both fast and slow noise is added according to an accurate noise model (Appendix \ref{app:noise-model-detail}).  Relaxation and decoherence times for the qubits are used during measurements and idling steps. Initialization and  measurement errors are modeled by a bit-flip channel preceding the measurements, that is, they are jointly subsumed under readout error. 

\subsection{Dephasing vs. readout: parameter exploration}
\label{sec:T2_vs_readout}

The goal of our proposal is to run the device in Fig.~\ref{fig:chip-layout} at increasing code distances and gather statistics on the logical error rates in order to test whether current technology is good enough to achieve error suppression in a spin qubit experiment, i.e. the device can operate below the error correction threshold of the code at hand. 

Looking at the circuit components, we first estimate the error rate of the involved single and two-qubit gates, which we do by averaging the channel over many ($10^3$) quasistatic noise realizations and then turning this channel into a Pauli channel by averaging over the Pauli group, which is known as the Pauli twirling approximation (PTA). For a channel given by ${\cal E}(\rho)=\sum_{m,n} \chi_{mn} P_m \rho P_n$ where $P_m$ are Paulis and $P_0=I$, the PTA takes the approximation ${\cal E}_{\rm PTA}=\sum_m \chi_{mm} P_m \rho P_m$ with $1-\chi_{00}$ the corresponding error rate. The gate fidelity is experimentally accessible through randomized benchmarking and is also the metric for which the pulse sequences are optimized, i.e. we list both $1-\chi_{00}$ as well as the gate (in-)fidelities computed by the pulse sequence finder designed by one of us (PC) in a different related project \cite{Cerfontaine2019a} in Table \ref{tab:gatefid} \footnote{Note that the infidelity components do not quite add up to the total infidelity in Table \ref{tab:gatefid}. We overestimate by concatenating the noise processes, because fidelity is not a linear function and averaging over slow and fast noise together is not necessarily the same as averaging over them separately and concatenating the two processes. Using averaged noise channels for fast gates saves computational effort, the price is that we are potentially overestimating the noise compared to joint averaging.}. We note that these fidelities and gate error estimates do not enter in our simulations as we simulate full gate dynamics, but the numbers are meant to provide guidance of expected performance and check  whether gate noise is biased towards particular Pauli errors. We have observed that the other entries on the diagonal of the $\chi$-matrix for the gates are not very biased towards a particular error with entries not varying over more than two orders of magnitude. 

\begin{table}[htb!]
\begin{center}
\begin{tabular}{c|c|c}
   & CNOT & $R_Y(\pi/2)$ \\
  \hline
  $\mathcal{F}$                      & $ 99.74\%           $  & $ 99.89\%             $ \\
  $1-\mathcal{F}_{\mathrm{slow}}$    & $ 9\times 10^{-4}   $  & $  1.5 \times 10^{-4} $ \\
  $1-\mathcal{F}_{\mathrm{fast}}$    & $ 1.9 \times 10^{-3}$  & $  1 \times 10^{-3}   $ \\
  $ p \approx 1-\chi_{00}$                        & $2.8\times 10^{-3}$              & $1.2 \times 10^{-3}$               \\
\label{tab:gatefid}
\end{tabular}

\caption{Gate metrics for the gate sequences we use in this work. The (in-)fidelities are the entanglement fidelities computed by the pulse sequence optimization developed in separate work by one of us (PC), see related publication \cite{Cerfontaine2019a}  The $\chi$-matrix was computed from these gate sequences as part of the present work in order to estimate the error rates of the gates. As a rough estimate we give the error rate as $1-\chi_{00}$.}
\end{center}
\end{table}

\begin{figure}[tb!]
\center{\includegraphics[width=1.1\columnwidth]
{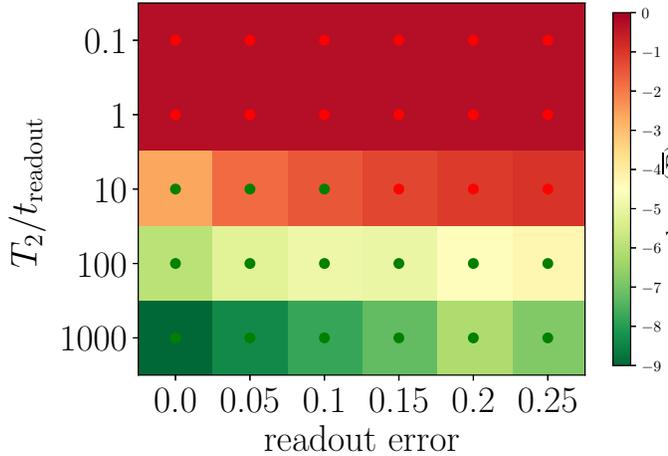}}
\caption{\label{phasediagram_T2_vs_readout} Phase diagram of the error enhancement to error suppression transition of the logical $Z$ error of the phase-flip code for lowering readout error ($x$-axis) and lowering qubit errors ($y$-axis), using a MWPM decoder. The criterion of error enhancement (red dots) and error suppression (green dots) is determined by a full circuit simulation of the real phase-flip code circuit with noisy components. This strongly indicates the feasibility of showing error suppression in the spin qubit device depicted in Fig. \ref{fig:chip-layout}. The logical error rate is shown as a heatmap (corresponding to the distance-$9$ code at each parameter tuple).}
\end{figure}

Comparing these numbers to the thresholds we found in Fig. \ref{depolarizing_logical_error_rate}, it is reasonable to believe that gate operations are below threshold for the phase-flip code. As described above, one of the bottlenecks at least in current and near term implementations of spin qubits is their long measurement times. While waiting for the measurement, the data qubits are idling, which as a noise process is strongly biased towards dephasing. Thus $T_2$ and $T_2^*$ are relevant figures of merit. We therefore undertake a numerical parameter study for a range of realistic estimates for integration time and readout fidelities. We explore the range between bare $T_2^*$ times and the most optimistic time-scales for echoing, by choosing $T_2/t_{\mathrm{readout}} = \{0.1,1,10,100,1000\}$, since we deem it reasonable to believe that the multi-dot device will be able to operate within this range (cf. \ref{sec:array}). The readout time $t_{\mathrm{readout}}$ is yet another parameter which is to be determined, however it is typically on the order of $1$ \textmu s, in which case the $y$-axis of Fig. \ref{phasediagram_T2_vs_readout} and related figures directly correspond to $T_2$-times. 

Since the readout fidelity is still to be characterized to high accuracy, we let this parameter range over a wide, possibly pessimistic, spectrum by allowing up to $25\%$ readout error. For every parameter pair we construct the circuit for several code distances $d=3,5,7,9$ and take $10^3 -10^4$ samples, which we decode with both the maximum likelihood decoder and a minimum weight decoder, where we adapt the weights by choosing a spatial weight $w_s=-\log(p)$ with the error rate given by the dephasing time and a temporal weight $w_t = -\log(q)$ with the measurement error rate $q$. For a code of distance $d$ we execute $d$ QEC cycles as in Section \ref{sec:decoding}. We observe that this MWPM decodes close to optimal (MLD) for our parameter ranges, but we do not plot the MLD data here.

For every parameter tuple, we determine whether the logical error rate is monotonically decreasing with increasing code distance $d$. If this criterion is fulfilled, we mark the tuple with a green dot, if it is not fulfilled, we mark it in red. Furthermore, to estimate the code performance we overlay the plot with a heatmap of the logical error rate of the distance-9 code at the corresponding parameter value. The result is shown in Fig. \ref{phasediagram_T2_vs_readout}. We observe that for $T_2 =10$ \textmu s with a given measurement integration time of $1$ \textmu s, the code could tolerate a readout infidelity of up to $15\%$. From the next order of magnitude in $T_2>100$ \textmu s 
the study suggests an extremely high tolerance to readout errors. This can be understood by looking back at the phenomenological anisotropic noise model, see Fig.~\ref{anisotropic_RBIM_phasediagram}. While the noise in our simulation is in principle not that simple since it occurs at the circuit-level and is spatially and temporally-correlated, this suggests that when the gates have low error rates we can indeed tolerate very high readout noise.

\subsection{The effect of increasing gate noise}
\begin{figure}[tb!]
\center{\includegraphics[width=1.1\columnwidth]
{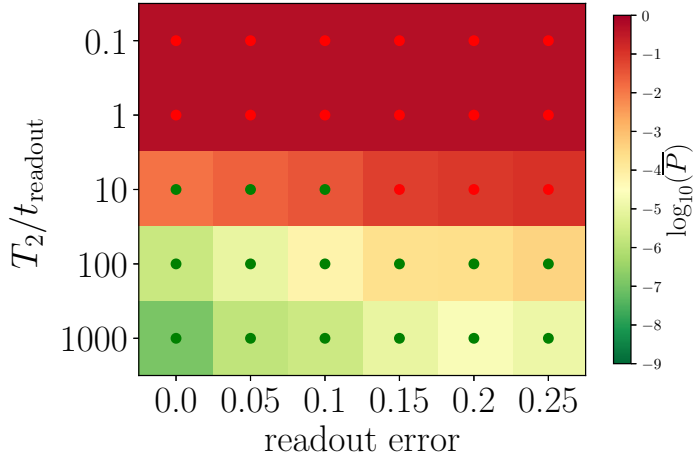}}
\caption{Cf. Fig.~\ref{phasediagram_T2_vs_readout}: Here we doubled the noise distribution width $\tilde \sigma_{\delta b,\delta \epsilon} =2\sigma_{\delta b,\delta \epsilon}$. The error suppression region slightly shrinks and the logical error rates increase by one order of magnitude, but the essential structure is maintained: the phase-flip code provides error suppression of phase noise for a wide range of parameters.}
\label{phasediagram_T2_vs_readout_double_noise}
\end{figure}

\begin{figure}[tb!]
\center{\includegraphics[width=1.1\columnwidth]
{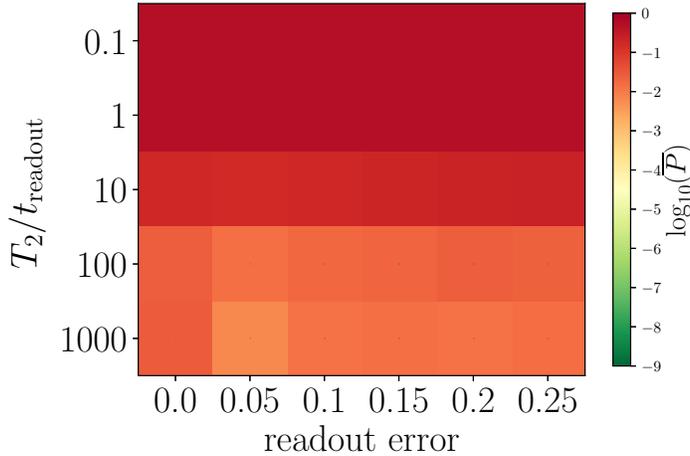}}
\caption{Cf. Fig.~\ref{phasediagram_T2_vs_readout}: Here we increase the noise distribution width fourfold: $\tilde \sigma_{\delta b,\delta\epsilon} =4\sigma_{\delta b, \delta \epsilon}$. The logical error rate increases drastically compared to the previous values of the noise distribution and one is nowhere below threshold.}
\label{phasediagram_T2_vs_readout_fourfold_noise}
\end{figure}

 In our gate simulations so far, we took the experimentally-measured values for the width of the distribution of slow $b$-field $\sigma_{\delta b}$ and exchange noise $\sigma_{\delta \epsilon}$ (Appendix \ref{app:noise-model-detail}). Since this noise has to this point never been characterized in a device of more than four dots, extrapolating these values might be too optimistic. To safeguard against possible increases in the noise distribution, we study the effect of a wider distribution by performing the simulation described in the previous paragraph \ref{sec:T2_vs_readout} with standard deviation of the noise distribution which is $2\times$ and $4\times$ as large for both the charge noise as well as the magnetic field fluctuations. The results can be seen in Figs. \ref{phasediagram_T2_vs_readout_double_noise} and \ref{phasediagram_T2_vs_readout_fourfold_noise}. For the latter, we do not put the error-enhancement vs. error-suppressing points since the logical error rates are substantially away from zero. We conclude that there is some leeway in the noise distribution, but we cannot operate at arbitrary parameter ranges.

\subsection{Coherence of noise and Pauli twirling}
\label{sec:PTA+coherence}
The question we want to address in this section is whether the coherence of the noise and the temporal and spatial correlations of slow noise are relevant in our numerics. The impact of applying a Pauli twirl approximation converting a general noise channel to a depolarizing noise channel and enabling scalable numerical simulations of Clifford circuits, instead of full-density matrix simulations, has already been studied in the literature, see e.g. \cite{Bravyi2018}. Results on small codes suggest that Pauli twirling overestimates the logical error rate, thus providing to some extent an upper bound on the logical error rate \cite{Katabarwa2015}. It has also been reported that Pauli twirling is a good approximation for incoherent noise models and worse for coherent errors \cite{Gutierrez2016}.

Since we use density-matrix simulations we can compare full noise simulations with those in which we remove both slow temporal correlations in the noise as well as applying the PTA approximation for each gate in the circuit. We compare the full circuit simulation reported above with a simulation where we replace every noisy gate by its perfect incarnation followed by an asymmetric depolarizing channel, which is obtained by twirling the true noisy gate error channel (sampled sufficiently often so the channel has a convergent representation). 

\begin{figure}[!tb]
\center{\includegraphics[width=1.1\columnwidth]
{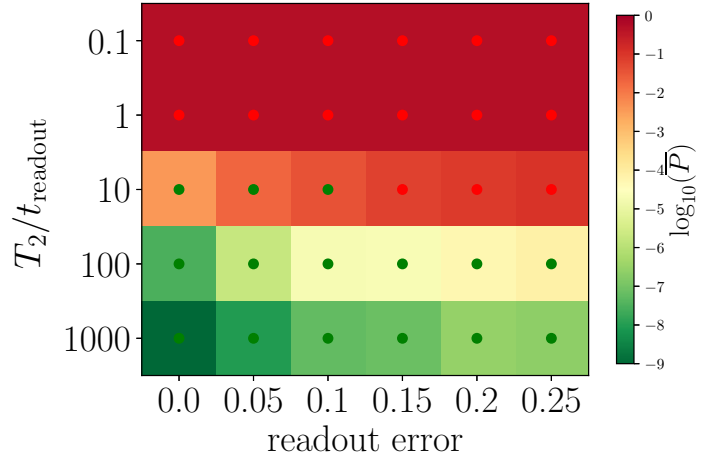}}
\caption{\label{phasediagram_T2_vs_readout_twirled} The effect of Pauli twirling: Analogous to Fig. \ref{phasediagram_T2_vs_readout}, we plot the phase diagram of the error enhancement to error suppression transition of the phase-flip code (using MWPM decoding), but here instead of simulating the full noise dynamics, we replace every noise process by its (slow and fast)-noise averaged-and-then-twirled version. }
\end{figure}

\begin{figure}[!tb]
\center{\includegraphics[width=1.1\columnwidth]
{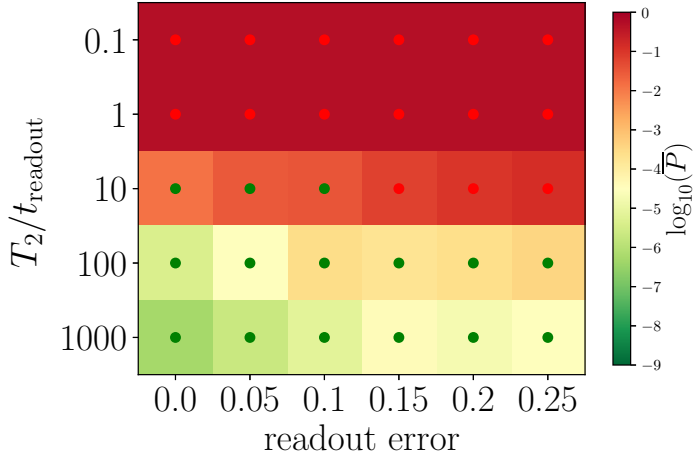}}
\caption{The effect of Pauli twirling: Analogous to Fig. \ref{phasediagram_T2_vs_readout_double_noise}, we increase the noise strength of slow noise by a factor of two in the distribution width $\tilde \sigma_{\delta b,\delta \epsilon} =2\sigma_{\delta b,\delta \epsilon}$ and plot the phase diagram of the error enhancement to error suppression transition of the phase-flip code, but here instead of simulating the full noise dynamics, we replace every noise process by its twirled and noise-averaged version.}
  \label{phasediagram_T2_vs_readout_twirled_double_noise}
\end{figure}

As shown in Figs. \ref{phasediagram_T2_vs_readout_twirled} and \ref{phasediagram_T2_vs_readout_twirled_double_noise}, the shape of the phase boundary stays the same and also the logical error rate differs only slightly. Our numerical results thus suggest that Pauli twirling does not dramatically alter the predictions of simulating a small phase-flip code in our regime of parameters. This has several explanations, probably the dominating noise is not gate noise, which is quite far below threshold in itself, no matter which error model. Furthermore, it is also not completely coherent, the contribution from slow noise is comparable to the contribution from fast noise. All in all we conclude that Pauli twirling seems to be a very acceptable numerical approximation in our regime, showing that costly full-density matrix simulations are in fact not needed to estimate the logical code performance for our parameter regime.


\subsection{The enhanced logical $X$-error rate}
\label{sec:enhanc}

Of course, dealing with a phase-flip code, we have no logical protection against $X$-errors. While the qubit noise is very biased towards phase-flips, the gates in the parity check circuit invariably introduce $X$-errors. 
If we were to start the QEC cycles in the state $\ket{\overline{0}}$ and finally destructively measure $\overline{Z}$, it gives us an estimate for the logical $\overline{X}$ error rate. We can roughly estimate this error rate by computing the probability that there are an odd number of $X$ errors on the final qubits (including errors in the final measurement) after $d$ QEC cycles. Let $p_{{\rm eff},x}$ be the effective $X$-error of a qubit per QEC cycle. This error rate is induced by the CNOT gate and $X$ errors induced by $T_1$- relaxation during the measurement time. Note that bit-flip errors coming out of a $R_Y(\pm \pi/2)$ gate do not enter the $X$-error rate, since they propagate to an even number of data qubits. In total, the logical qubit suffers an $X$-error if there are an odd number of $X$-errors in a space-time volume of size $d^2$ (composed of $d$ QEC cycles on $d$ qubits), i.e. 
\begin{align}
 \overline{P}_{X}(p_{{\rm eff},x}) &= \sum_{k\;\;\mathrm{odd}} {d^2 \choose k} p_{{\rm eff},x}^k (1-p_{{\rm eff},x})^{d^2-k} \notag \\
  &= \frac{1}{2}\left(1-(1-2p_{{\rm eff},x})^{d^2}\right)
\label{eq:eff}
\end{align}
Neglecting the effect of $T_1$-errors, we estimate that the CNOT introduces an $X$ error on the data qubit with error rate $p_{{\rm eff},x} \approx 0.0013$. For comparison, we compute the logical $X$-error rate in the circuit simulation by sending the state $\ket{\overline{0}}=\frac{1}{\sqrt{2}}(\ket{++\ldots +}+\ket{--\ldots -})$ through the circuit. The relative sign of this state is flipped by $X$-errors and thus we can determine the logical $X$-error rate by determining the probability that the final state has negative relative sign. The values of this are in good agreement with the phenomenological formula in Eq.~\ref{eq:eff} at $p_{{\rm eff},x} \approx 0.0013$.

\begin{figure}[tb!]
\center{\includegraphics[width=1\columnwidth]
{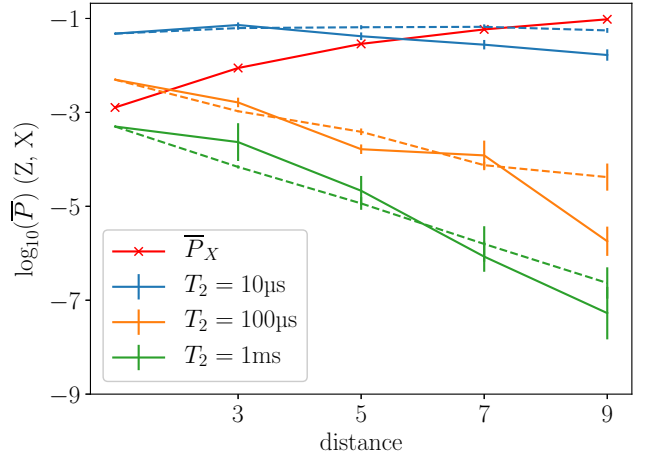}}
\caption{\label{logical_noise_tradeoff} Logical error rate tradeoff of the phase-flip code. Solid $\overline{P}_Z$ is for readout error $5\%$ and dashed for $15\%$. For sufficiently-long $T_2$ times (with $t_{\rm readout}=1$\textmu s), by encoding in the phase-flip code, the logical $Z$-error rate is exponentially suppressed. In contrast to this, the logical $X$-error rate increases with larger encoding circuit. The distance-3 code has a logical $X$-error rate at the value of the surface code threshold, which upon noise optimization could allow for a sub-threshold error rate for a code concatenation.}
\label{fig:tradeoff}
\end{figure}

We show the behavior of the logical $X$-error rate in Eq.~(\ref{eq:eff}) for increasing code distance compared to the logical $Z$-error rate for the same distance, which we plot for two settings of readout error ($5\%$ and $15\%$) in Figure \ref{logical_noise_tradeoff}. As expected, the logical $X$-error is growing for large distances, however the distance-$3$ code is not hopeless. It sits close to the surface code threshold at $1\%$ \cite{Wang2003}. Note that the logical $X$-error is extremely sensitive to the $X$-error rate of the CNOT gate. The gate sequences could be taylored to specifically avoid these types of errors by adapting the target function of the pulse sequence optimization routine, thereby potentially pushing the logical $X$-error rate below surface code threshold, see also Section \ref{sec:outlook}.

\subsection{Leakage and leakage reduction}
\label{sec:leakage}
The qubits employed in the envisioned architecture are encoded in the $\{|S\rangle,|T_0\rangle\}$-subspace of two electrons in a double dot. The remaining triplet states $\{|T_+\rangle,|T_-\rangle\}$ comprise non-computational leakage states and the most relevant process for leakage in our setup are faulty gate operations \cite{Cerfontaine2014}. Continued leakage accumulation poses a threat to error correction \cite{AT:leak, Fowler2013}, but the problem can be addressed using so-called Leakage Reduction Units (LRUs). These units return leaked qubits to the computational subspace, potentially introducing qubit errors, which can be handled by the error correcting code \cite{Mehl2015}. For singlet-triplet qubits, such a LRU can be implemented by a Swap If Leaked (SIL) procedure \cite{Mehl2015}, swapping the concerned (data-)qubit with a fresh ancilla qubit only in case the data qubit is in a leakage state. A compact way of representing the desired operation (see Section \ref{sec:qubits} for qubit and leakage states) is a truth table specifying the required mapping for such a SIL procedure \cite{Mehl2015}:
\begin{alignat*}{3}
&|S^\mathrm{D}\Psi^\mathrm{A}_0\rangle&&\rightarrow&&|S^\mathrm{D}\Psi^\mathrm{A}_1\rangle,\\
&|T_0^\mathrm{D}\Psi^\mathrm{A}_0\rangle&&\rightarrow&&|T_0^\mathrm{D}\Psi^\mathrm{A}_1\rangle,\\
&|T_-^\mathrm{D}\Psi^\mathrm{A}_0\rangle&&\rightarrow&&|\Psi^\mathrm{D}_\mathrm{r,-}T_-^\mathrm{A}\rangle,\\
&|T_+^\mathrm{D}\Psi^\mathrm{A}_0\rangle&&\rightarrow&&|\Psi^\mathrm{D}_\mathrm{r,+}T_+^\mathrm{A}\rangle,
\end{alignat*}
where $|\Psi^\mathrm{A}_0\rangle$ and $|\Psi^\mathrm{A}_1\rangle$ are initialization and output states in the computational ancilla subspace $\{|S^\mathrm{A}\rangle,|T_0^\mathrm{A}\rangle\}$ and $|\Psi^\mathrm{D}_\mathrm{r,-}\rangle$ and $|\Psi^\mathrm{D}_\mathrm{r,+}\rangle$ are reset states in the $\{|S^\mathrm{D}\rangle,|T_0^\mathrm{D}\rangle\}$-subspace, all of which may be chosen arbitrarily. We have used D for data qubit and A for ancilla qubit, i.e. we are swapping leakage from data to ancilla qubit. This gadget is a two qubit interaction, which can be implemented analogously to the two-qubit gates we employ elsewhere in this work, so that it is reasonable to assume that the SIL sequence can be similar in both duration ($50\,\mathrm{ns}$ cf. Table \ref{tab:times}) as well as in fidelity. A neat feature of the proposed chip design (Fig. \ref{fig:chip-layout}) is that it allows for fast energy selective reset into a singlet state via coupling to a nearby electron reservoir,  such that it is possible to execute the leakage reduction sequence including ancilla reset on a time-scale of tens to hundreds of nanoseconds depending on the initialization method of $|\Psi^\mathrm{A}_0\rangle$ (see Section \ref{sec:initialization}).

\begin{figure}[tb!]
	$$
	\begin{tikzcd}
	& \gate{\mathcal{E}_{\mathrm{leak}}} & [2mm]\gate[wires=2]{\mathrm{LRU}} & [2mm]\qw \\
	\lstick{$\ket{\Psi^\mathrm{A}_0}$} & \qw & & \qw\rstick{$\skull$} \\
	\end{tikzcd}
	\equiv
	\begin{tikzcd}
	& [2mm]\gate[wires=1]{\mathrm{\mathcal{E}_\text{eff}}} & [2mm]\qw\\
	& & \\
	\end{tikzcd}
	$$
        \caption{The error channel of a data qubit undergoing a leakage process is converted into an effective qubit channel by using an ancilla that is swapped in by the LRU gate in case of leakage. The ancilla qubit is afterwards discarded (or reset and used as a ancilla qubit for QEC), which we represent by the symbol $\skull$.}
        \label{fig:leakage_reduction_channel}
      \end{figure}
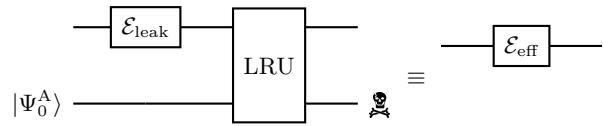

\begin{figure}[tb!]
	\begin{tikzcd} 
		\lstick{D}& \gate[]{\mathcal{E}_\mathrm{Echo}}\gategroup[wires=6,steps=1,style={inner sep=2.5pt}]{ME} &\qw & \gate[wires=2]{\mathrm{LRU}}\gategroup[wires=7,steps=1,style={inner sep=2.5pt}]{LR} &\qw& \gate[wires=6]{\mathrm{SG}}&\qw\\
		\lstick{A}& \meter{} &\push{\ket{\Psi^\mathrm{A}_0}} && \push{\skull} & & \qw\\
		\lstick{D}& \gate[]{\mathcal{E}_\mathrm{Echo}} &\qw &\gate[wires=2]{\mathrm{LRU}} &\qw& &\qw\\
		\lstick{A}& \meter{} &\push{\ket{\Psi^\mathrm{A}_0}} && \push{\skull} & & \qw\\
		&\gate{\qquad}\qwbundle[alternate]{}
		&\qwbundle[alternate]{}
		&\gate{\qquad}\qwbundle[alternate]{}
		&\qwbundle[alternate]{}
		&\qwbundle[alternate]{}
		&\qwbundle[alternate]{}
		\\
		\lstick{D}& \gate[]{\mathcal{E}_\mathrm{Echo}} & \qw & \gate[wires=2]{\mathrm{LRU}} &\qw& &\qw\\
		\lstick{A}& \qw  & \push{\ket{\Psi^\mathrm{A}_0}} & &  \push{\skull} & \qw & \qw
	\end{tikzcd}
	\caption{Circuit arrangement of the operation blocks to sustain the logical qubit including leakage reduction. During the Measurement and Echo (ME) block (of duration $>1\,$\textmu s) syndromes are collected from the ancillas, while the data qubits are subject to echo sequences, which are assumed to be the main culprit for inducing leakage (see text). The subsequent Leakage Reduction (LR) converts the leakage into qubit errors in the computational subspace, which are passed to the following Stabilizer Gate (SG) block where the two qubit interactions take place. As each data qubit has to be paired up with an ancilla partner during the LR block, one extra ancilla qubit is added to the array.}
        \label{fig:leak-in-circ}
      \end{figure}
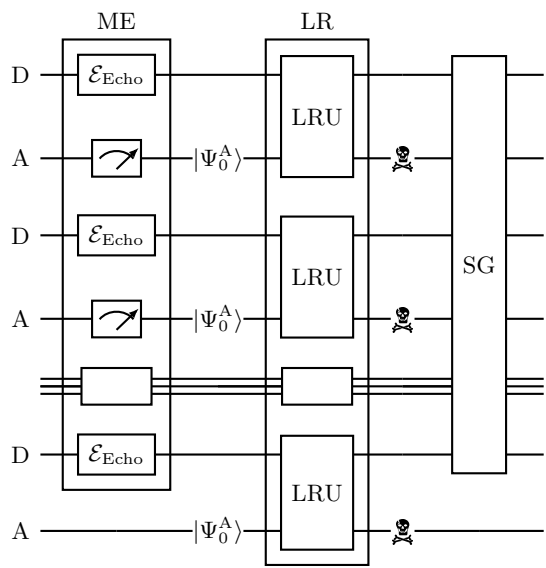

For the chosen qubit type, material system and architecture, leakage induction will be dominantly due to execution of gates (with e.g. the single qubit gate leakage rate measured to be 0.13\% \cite{Cerfontaine2019b}). Therefore, most of the leakage will be accumulated during the echoing sequence executed on the data qubits while the ancilla qubits are being measured. These echoing techniques are necessary to generate the high effective $T_2$ used throughout the paper, but need of the order of 10 single qubit gates to reach $T_2$ times that ensure sufficient qubit coherence after the measurement time has passed (e.g. in \cite{Bluhm2010b} 16 gates were needed to reach $200$\textmu s). We therefore place the leakage reduction block directly before the stabilizer gates (SG, two qubit interactions) between data and ancilla qubits (see Fig. \ref{fig:leak-in-circ}), allowing to remove leakage acquired during the echoing sequence from the data qubits before reaching the stabilizer block. Conversion of leakage to errors in the computational subspace allows for the error correcting circuit to detect and correct them. This is essential, since having leaked states as input to the two-qubit gates of the parity checks can cause checks to flip or spread errors (depending on experimentally and theoretically unexplored details). Performing an LRU operation has the effect that a leaked data qubit state is re-initialized in the computational qubit subspace, which from a quantum information perspective means that we can effectively ignore the ancilla (tracing it out) and can describe the effective single \emph{qubit} channel (emphasis on qubit, i.e. not leakage), as shown in Fig. \ref{fig:leakage_reduction_channel}.

Let us describe the leakage process by some incoming leakage error rate $p_{\mathrm{leak}}$ with a corresponding channel
\begin{equation}
\mathcal{E}_{\mathrm{leak}}(\rho) = (1-p_{\mathrm{leak}})\rho + p_{\mathrm{leak}} \Sigma,
\end{equation}
where $\rho$ is some qubit state, but importantly $\Sigma$ is some state in the leakage space, i.e. some mixture of $\ket{T_+}$ and $\ket{T_-}$. The effect of the LRU is to transform $\Sigma$ into a reset state inside the computational subspace. This means that the leakage channel is transformed into the effective channel
\begin{equation}
\mathcal{E}_{\mathrm{eff}}(\rho) = \mathcal{E}_{\mathrm{LRU}}\left( \mathcal{E}_{\mathrm{leak}}(\rho)\right) = (1-p_{\mathrm{leak}}) \rho + p_{\mathrm{leak}} \rho_{\rm reset},
\end{equation}
where $\rho_{\rm reset}$ is some state supported in the space spanned by $\ket{T_0}$ and $\ket{S}$. Here we are ignoring higher order contributions, such as the LRU itself being a noisy process. 
If this error channel produces $Z$ errors, they can be detected by subsequent QEC cycles of the phase-flip code. We therefore propose to perform a leakage reduction procedure directly before the error correction unit, as shown in Fig.~\ref{fig:leak-in-circ}.
It should be noted that both the effective coherence time $T_2$ during the echo sequence as well as the accumulated leakage in the data qubits increase with the number of gates ($\pi$-pulses) executed during the echo sequence, as described in the previous paragraph. This implies that it might be a bit naive to strive for $T_2$ times as large as possible, since this directly corresponds to increased leakage rates. Furthermore, the larger the number of gates in the echo sequence, the more errors of the Pauli-$X$ type accumulate (see Section \ref{sec:enhanc}), which are not detectable by the QEC code at hand. This again implies a trade-off that probably makes very high $T_2$ times less desirable than they might seem at first glance.
\section{Discussion and outlook: towards the surface code?}
\label{sec:outlook}
Here we discuss several open questions relating to the feasibility of moving beyond a small phase-flip code.

A layout based on a linear chain of quantum dots is mostly likely not easily scalable to the large number of qubits required for building a practically useful quantum computer.
While we mitigated the fan-out problem by introducing extra gate layers, this issue will reappear as more qubits are added. In addition, connecting a large number of gates to off-the-shelve room temperature electronics via dedicated wires is also a serious obstacle for scaling to large numbers of of qubits. In the near future, both of these problems could be solved by integrated control electronics which can be connected to the qubit chip via flip-chip bonding, thus reducing the fan-out and the need for wires \cite{Geck2019}.

In addition to issues with wiring and fan-out, control cross-talk will complicate the operation of closely packed qubits. Since true two-qubit cross-talk caused by unwanted two-qubit interactions is likely not an issue for singlet-triplet qubits \cite{Cerfontaine2019a}, compensating linear control cross-talk between adjacent control lines and qubits does not present a fundamental obstacle. However, calibrating cross-talk in large arrays of tightly packed qubits could still prove difficult if the cross-talk does not fall off quickly enough with the distance between qubits. 

In addition, it could be necessary to consider the effects of second order tunneling or mediated exchange \cite{Mehl2014,Baart2016c} if the exchange interaction between several dots is turned on simultaneously. Furthermore, DNP \cite{Bluhm2010} has not been realized in large quantum dot arrays so far, but should also not pose a fundamental obstacle. However, coherence times are unlikely to be longer than demonstrated for isolated qubits as additional error mechanisms and decay channels are added.

If a 2D device architecture were available, one may ask whether the use of the phase-flip code would still be of interest. For example, one could directly use the surface code with limited `cross-bar' control as envisioned in \cite{crossbar} or consider faster, measurement-free, use of small surface codes \cite{ercan+:meas-free}. 
One could imagine using a clustered 2D layout with each cluster representing a logical qubit encoded by the phase-flip code connected by long-range CNOTs with other clusters or ancilla qubit, while short-range CNOT gates are used inside the cluster.

The answer to this `best architecture' question cannot fully be given at this stage and depends on various considerations, for example to what extent the noise is biased towards $Z$-errors. Let us give some considerations.

For GaAs singlet-triplet qubits, even when performing dynamical decoupling, a factor of 10 between the $X$-error rate and $Z$-error rate may be expected. Similarly, idling Si-based qubits have a strong noise bias towards $T_2$, see Table \ref{tab:si}. If two-qubit gates are of very high fidelity (with negliglible error) and noise is dominated by phase-flip errors during slow, noisy measurements, then, instead of using the phase-flip code as bottom code, one might be able to directly benefit from using a modified surface code in which one measures $X$- and $Y$-checks, considered in \cite{tuckett:bias}. With an (incoming) phenomenological $Z$-error rate $p_Z \approx p$ and $X$- and $Y$-error rate equal to $p_{X}=p_Y \approx p/200$ (noise bias $\eta=100$) and a measurement error rate equal to $p$, the reported threshold in \cite{tuckett:bias} is $p_c=5\%$. For the GaAs numbers in Tables I and II, $p_Z \approx t_{\rm readout}/T_2=1.25 \times 10^{-3}$ and $p_X \approx t_{\rm readout}/T_1=5 \times 10^{-4}$, putting one safely below this $5\%$ threshold.

However, this picture is not realistic since the two-qubit gate error rates are in fact not negliglible and biased-noise or bias-preserving gates have not been the focus of previous research. For example, doing 4 CNOTs in a surface code QEC cycle, each CNOT gate would have to have $X$-error rate at least below $\frac{5\%}{800}=0.625 \times 10^{-4}$ in order to get below this $5\%$-threshold of \cite{tuckett:bias}. It might be interesting to develop such noise-bias preserving CNOT with dominant $Z$-noise as has been done for superconducting devices in \cite{puri+:bias}.

Given that the noise-bias is unlikely to be a direct advantage in surface code decoding a la \cite{tuckett:bias}, we can further consider using the phase-flip code as the bottom code to equalize the $X$- and $Z$-error rates as we have numerically explored in Section \ref{sec:enhanc}. Thus the phase-flip repetition code could then be concatenated, in principle, with the surface code.

\begin{figure}[tb!]
\begin{tikzcd} 
& [5mm]\gate{\rm QEC}\qwbundle{3} &\qw & \targ{} & \qw & \qw & \qw & \gate{\rm QEC} & \qw \\
& \gate{\rm QEC}\qwbundle{3} &\qw & \qw & \targ{} & \qw & \qw & \gate{\rm QEC} & \qw \\
& \gate{\rm QEC}\qwbundle{3} & \qw & \qw & \qw & \targ{} & \qw &  \gate{\rm QEC} & \qw \\
& \gate{\rm QEC}\qwbundle{3} & \qw & \qw & \qw & \qw & \targ{} &  \gate{\rm QEC} & \qw \\
\lstick{$\ket{+}$} &\qw & \qw & \ctrl{-4} & \ctrl{-3} & \ctrl{-2} & \ctrl{-1} & \meter{$X$} & \\
\end{tikzcd}
\caption{Surface code $X$-parity check circuit, using data qubits each encoded with the 3-bit phase-flip code suggested by the number 3 on the lines. Each QEC-unit denotes 3 or more cycles of phase-flip error correction and is repeated continuously, interspersed by CNOT gates. The surface code ancilla could be chosen to be encoded, so that $\ket{+}$ denotes a logical $\ket{\overline{+}}$ and $X$ denotes the logical $\overline{X}$ measurement for this ancilla. The preparation of $\ket{\overline{+}}=\ket{+++}$ is simple and the logical $\overline{X}$ measurement is robust by majority-vote taking. However, since the circuit is short in duration, using a single-qubit prepared in $\ket{+}$ as ancilla will most likely have better performance. 
Note that each CNOT in the circuit has to be done transversally, between triples of qubits when the ancilla is encoded, or between a triple and a single ancilla qubit when the ancilla qubit is unencoded.
}
\label{fig:XXXX}
\end{figure}
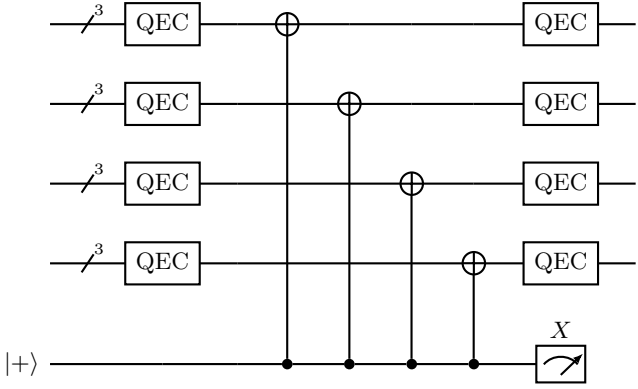

\begin{figure}[tb!]
\begin{tikzcd}
& [5mm]\gate{\rm QEC}\qwbundle{3} & \qw & \ctrl{4} & \qw & \qw & \qw & \gate{\rm QEC} & \qw \\
& \gate{\rm QEC}\qwbundle{3} & \qw & \qw & \ctrl{3} & \qw & \qw & \gate{\rm QEC} & \qw \\
& \gate{\rm QEC}\qwbundle{3} & \qw & \qw & \qw & \ctrl{2} & \qw & \gate{\rm QEC} & \qw \\
& \gate{\rm QEC}\qwbundle{3} & \qw & \qw & \qw & \qw & \ctrl{1} & \gate{\rm QEC}& \qw \\
\lstick{$\ket{0}$} & \qw & \qw & \targ{} & \targ{} & \targ{} & \targ{} & \meter{$Z$} & \\
\end{tikzcd}
\caption{Surface code $Z$-parity check circuit on phase-flip encoded triples of qubits. The ancilla could be replaced by its phase-flip encoded version, but the logical $\overline{Z}$ measurement is sensitive to measurement errors and the state $\ket{\overline{0}}$ is not simple to prepare. It is thus certainly better to use a single ancilla qubit started in $\ket{0}$ and a direct $Z$-measurement.}
\label{fig:ZZZZ}
\end{figure}
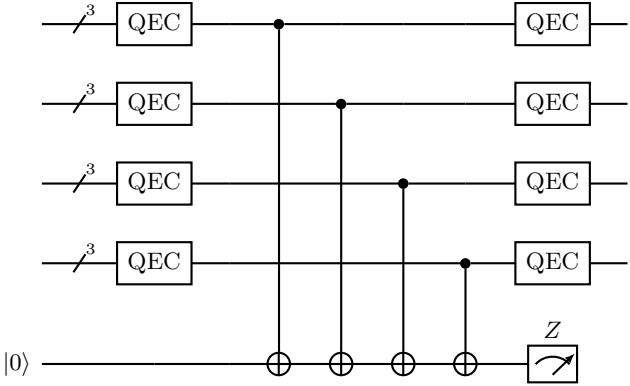

If a strongly biased-noise CZ gate (with dominant $Z$ errors) was available for spin qubits, one could imagine following an approach advocated in \cite{Aliferis2009} with the phase-flip code as bottom code. Note however that the logical CNOT gadget for the phase-flip code from \cite{AP:bias} which aims to preserve this noise-bias is non-ideal as it heavily uses slow qubit measurements and additional qubit overhead. So we won't focus on obtaining a CNOT in this way. Let us now discuss the possible drawbacks and variations of this concatenation scheme.

First, let us imagine encoding each qubit of the surface code in the phase-flip code, including surface code ancilla qubits. One then imagines alternating 3 or more cycles of phase-flip error correction with applying logical gates, e.g. the CNOT gates to implement the parity check circuits of the surface code as in Fig.~\ref{fig:XXXX} and \ref{fig:ZZZZ}. As the phase-flip repetition code is a CSS (Calderbank-Shor-Steane) code, the logical CNOT gate is a transversal gate, meaning that it can be realized by applying copies of the basic CNOT gate between two code blocks \footnote{One can understand this by tracking the XX-stabilizers through the CNOT as well as the logical operators.}.

Now consider the logical error rate induced by such transversal CNOT gate if we use the distance-3 phase-flip code. Since the phase-flip code has {\em no} correction against $X$ errors, we have a logical failure on one of the encoded qubits when any one of the 3 CNOT gates produces an outgoing $X$-error. Said differently, the use of the 3-qubit phase-flip code would make us suffer from a CNOT-induced error rate which is $3\times$ higher than if we were to use no phase-flip encoding but use the surface code `straight-up'. Given that the surface code cannot tolerate an error rate of more than $1\%$ of circuit-level depolarizing noise \cite{fowler+:surf}, such form of concatenation puts some very strong demands on the CNOT $X$-error rate. 

One simplification of this scheme is to use bare ancilla qubits to measure the toric code checks instead of phase-flip encoded ancilla qubits, see the circuits in Fig. \ref{fig:XXXX} and Fig.~\ref{fig:ZZZZ}. This is advantageous since these ancilla qubits are only short-lived so preparation and measurement errors are dominant sources of errors. In particular, the logical preparation of $\ket{\overline{0}}$ and logical measurement $\overline{Z}$ are cumbersome in the phase-flip code. If we consider Fig.~\ref{fig:ZZZZ}, a transversal CNOT gate between encoded data qubit and encoded ancilla then gets replaced by 3 CNOT gates between encoded data qubits and the single-qubit ancilla. These 3 CNOT gates propagate any $X$-error on the data qubits (a logical $\overline{X}$ for the phase-flip code) to the ancilla and makes it flip and thus detectable. If two out of the three qubits have an $X$-error, it is equivalent to a stabilizer and the ancilla will not flip.
For the $X$-parity circuit, even though one could use an encoded ancilla as the preparation of $\ket{\overline{+}}=\ket{+++}$ is straightforward and its measurement robust, its performance will be no better than using a simple $\ket{+}$ and measuring in the $X$-basis. So again, we apply 3 CNOTs between an encoded data qubit and the $\ket{+}$ ancilla. Any odd number of $Z$-errors on a data qubit, either the logical $\overline{Z}=Z_1 Z_2 Z_3$ or $Z_i$ will thus make the ancilla flip from $\ket{+}$ to $\ket{-}$. This has to be taken into account in the decoding of the toric code, i.e. information about dephasing errors on the phase-flip code has to be combined with toric code information. 

However, now that we consider using single uncorrected ancillas, one can also see the whole scheme as one big code. This code has local adjacent $XX$-checks on the clusters (due to the phase-flip code) which are measured continuously in the EC-units in Figs.~\ref{fig:XXXX} and \ref{fig:ZZZZ}. In addition, the code has (high) weight-12 $Z$-checks for surface code $X$-error correction as well as the usual weight-4 $X$-checks (or alternatively, weight-12 as well) for $Z$-error correction. It then seems that this scheme can only be of interest when short-range and long-range CNOT have extremely low error rates so that the gains of the phase-flip code are not undone by the losses involved in having to do many CNOT gates.

All-in-all, we see that the achievable quality and bias of the two-qubit gate is crucial in setting the optimal use of error correction beyond the scheme presented in this paper. Before these numbers are experimentally in place, it does not make a whole lot of sense to further speculate on this. 

\section{Acknowledgments}

Our work was supported by ERC grant EQEC No. 682726 and the Impulse and Networking Fund of the Helmholtz Association. We thank Hendrik Bluhm, David DiVincenzo and Lieven Vandersypen for feedback and discussions. MR would like to thank Benedikt Placke for discussions on the statistical-mechanical mapping of decoders. PC would like to thank Matthias K\"unne for discussions about the device design. We acknowledge computing resources of RWTH Aachen under project rwth0274.
\appendix

\section{Details of the noise model used in numerical simulation}
\label{app:noise-model-detail}
Noisy gates can be modeled by perfect gates instantaneously followed by a noise channel describing the deviation of the actual noisy gate from the perfect (target-) gate. We model the single- and two-qubit gate similarly to how it was done in \cite{Cerfontaine2019a}, decomposing the noise into slow and fast components, where in our simulations we additionally account for the fact that we do not only care about a single gate but about the performance of many QEC cycles. As described in Section \ref{sec:operations}, the dominant noise processes can be understood as variations to the control signals that comprise the execution of the desired gate. To remind the reader, these are the dot detunings $\bm{\epsilon}$ and the magnetic field gradients $\bm{b}$, such that the relevant noise parameters are random fluctuations around these values, which we denote by $\delta \bm{\epsilon}$ and $\delta\bm{b}$. In case of a two-qubit gate these are thus three variables each, for all the exchange couplings and gradients between neighboring dots. The pulse sequence takes the time $t_\mathrm{{gate}}$, during which we modulate the detuning quickly. Let us now write this time-dependence explicitly as $\bm{\epsilon}(t)$. This is in principle a continuous analog signal, however to very good approximation described by a discretization into time-steps, i.e. a discrete time-series of detunings. Adding noise to this signal amounts to adding another time-series on top of the intended one for each individual $\epsilon_{ij}$ and $b_{ij}$ (the intended one comes out of a numerical optimization as also stated in Section \ref{sec:operations}). The noise signal is therefore a time-series that can be described by the methods of time-series analysis typically done in terms of the so-called power spectrum of the autocorrelation function of the signal (see e.g. \cite{Paladino2014}). This analysis describes the noise signal in the frequency domain, which lets us decompose the noise into fast and slow components with respect to the gate execution time. The slow part are variations to the signal that do not change over the duration of the gate, we thus model them as a random value set once and kept over the gate execution time. These random variables can be efficiently obtained by drawing from a normal distribution with a standard deviation that comes out of experimental observations (in our case $\sigma_{\delta \epsilon} = \SI{8}{\micro V}$ \cite{Dial2013} and $\sigma_{\delta b} = \SI{0.3}{mT}$ \cite{Cerfontaine2016}). That is, we draw as many uncorrelated random variables as there are detunings in the setup. This constitutes a noise instance of slow noise, which we can add to the detuning time sequence. The same analysis holds for the magnetic field gradients in complete analogy, moreover for magnetic field noise it turns out that the noise is dominated by slow components \cite{Reilly2008}, such that the above method is exhaustive. For charge noise on the detuning, we have to go further. Modeling the fast noise component is slightly more involved, since here we have to take the variations during the gate sequence duration into account. We generate noise instances of fast noise by imitating the power spectrum that has been characterized experimentally for our system (GaAs): the high-frequency spectrum follows $S_{\epsilon,\alpha}(f) \propto 1/f^\alpha$ with $\alpha = 0.7$ \cite{Dial2013}. Since this spectrum has not been measured above a few MHz, which are still relevant frequencies for our gates, we extrapolate with $\alpha = 0.7$ such that $S_0 = \SI{4e-20}{V^2/Hz}$ at \SI{1}{MHz} \cite{Dial2013}. With the ability to generate fast and slow noise instances, we now turn to how we obtain the effective noise channels that enter our simulations as described in Fig. \ref{fig:circuit_t2_vs_readout}. On the level of an individual noise instance, a noise signal just has to be added to the intended pulse sequence and magnetic field gradients before integrating the Schr\"odinger equation (Eq. \ref{time_evolution_gate}), which gives an instance of a unitary
\begin{equation}
\tilde{U} = \prod_{m=1}^{N} \exp(-iH(\bm{\epsilon}_m + \delta \bm{\epsilon}, \bm{b}+\delta \bm{b}) \Delta t).
\end{equation}
that is the noisy version of a perfect unitary $U$. In order to capture the noisy part of this gate, we undo the ideal part and give the noisy unitary part the name $E$:
\begin{equation}
  E = U^\dag \tilde{U}.
  \label{noise_unitary_instance} 
\end{equation}
Whether $E$ describes slow or fast noise only alters what we put into $\delta \bm{\epsilon}$ and $\delta \bm{b}$. For slow noise, we draw these parameters $\delta \bm{\epsilon}_\mathrm{slow}$ and $\delta \bm{b}_{\mathrm{slow}}$ once per realization of the whole QEC circuit, such that all gates during the circuit execution experience the same instance of slow noise (the same offset of the detuning and the magnetic field gradients). 

Since we work with a full density-matrix simulation, we express this as a quantum channel
\begin{equation}
\mathcal{N}_{\mathrm{slow}}(\rho) = E_{\mathrm{slow}} \rho E_{\mathrm{slow}}^\dag.
\end{equation}
In contrast to this, when we want to describe fast noise processes, we generate many $(\Omega = 10^4)$ realizations of the random signal $\delta \bm{\epsilon}_{\mathrm{fast}}$, each time integrating the Schr\"odinger equation to obtain a unitary instance $E_k$ of Eq.~\ref{noise_unitary_instance}, where the index runs over the number of realizations. We then describe the fast noise process as the average quantum channel by averaging over these instances, i.e.
\begin{equation}
\mathcal{N}_{\mathrm{fast}}(\rho) = \frac{1}{\Omega} \sum_{k=1}^{\Omega} E_k \rho E_k^\dag.
\end{equation}
The combined noise process $\mathcal{N}$ is then described by concatenating slow and fast noise:
\begin{equation}
\mathcal{N} = \mathcal{N}_{\mathrm{fast}} \circ \mathcal{N}_{\mathrm{slow}}.
\end{equation}
We implemented these channels for the gates needed for the phase-flip code (Fig.~\ref{fig:circuit_t2_vs_readout}) in Quantumsim. Amplitude-damping and phase-damping are described in the Kraus representation $\mathcal{E}(\rho) = \sum_i A_i\rho A_i^\dag$ by the following Kraus channels:

\begin{equation}
  A_0^{\mathrm{AD}}= \begin{pmatrix} 1 & 0\\ 0 & \sqrt{1-\gamma} \end{pmatrix} \quad   A_1^{\mathrm{AD}}= \begin{pmatrix} 0 & \sqrt{\gamma}\\ 0 & 0 \end{pmatrix},
\end{equation}

\begin{equation}
  A_0^{\mathrm{PD}}= \begin{pmatrix} 1 & 0\\ 0 & \sqrt{1-\eta} \end{pmatrix} \quad   A_1^{\mathrm{PD}}= \begin{pmatrix}0 & 0\\  0 & \sqrt{\eta} \end{pmatrix},
\end{equation}
where $\gamma$ and $\eta$ are the amplitude and phase damping rates. These rates are typically expressed as times $T_1$ (relaxation or amplitude damping time), $T_\varphi$ (pure dephasing time, phase-damping) and $T_2$ (decoherence time), for which it holds that $e^{-t/T_1} = 1-\gamma$, $e^{-t/T_\varphi} = \sqrt{(1-\eta)}$ and $e^{-t/T_2} = \sqrt{(1-\gamma)(1-\eta)}$. Since in our setup, $T_1\gg T_2$, decoherence acts like pure dephasing noise $T_2 \approx T_\varphi$, which as a noise channel can be also written (as seen e.g. by exploiting the unitary freedom in the Kraus representation) as a phase-flip channel ${\mathcal S}(\rho) = (1-p_Z) \rho + p_Z Z\rho Z$ with error rate
\begin{equation}
  p_Z = \frac{1-e^{-t/T_\varphi}}{2} \approx \frac{1-e^{-t/T_2}}{2}.
\end{equation}

\section{Zeeman Hamiltonian with magnetic field gradients}
\label{app:Zeeman}
For completeness we here show the Hamiltonian in Eq.~\ref{Qubit_Hamiltonian} expressed in terms of the magnetic field gradients. The Zeeman Hamiltonian is simply $H_{\mathrm{Zeeman}} = \frac{1}{2}\sum_i B_i \sigma_z^i$. The aim is a change of variables $\bm{B} \rightarrow (B_{\mathrm{G}},\bm{b})$ with $b_{ij} = B_j -B_i$ and $B_{\mathrm{G}}=\frac{1}{4}\sum_i B_i$. This implies a change-of-basis matrix
\begin{equation}
  R = \begin{pmatrix}
    \frac{1}{4} & \frac{1}{4} & \frac{1}{4}& \frac{1}{4}\\
    -1 & 1 & 0 & 0\\
    0 & -1 & 1 & 0 \\
    0 & 0 & -1 & 1 \\
    \end{pmatrix}.
  \end{equation}
  
Evidently the Pauli-$Z$ matrices have to be transformed with $\left(R^{-1}\right)^\mathrm{T}$, such that the total Hamiltonian in Eq.~\ref{Qubit_Hamiltonian} is expressed as
\begin{align}
  H(\bm{\epsilon},\bm{b}) = &\frac{1}{2}  B_\mathrm{G}  \sum_{i=1}^4 {\sigma_z}^{(i)} \nonumber \\
	+ & \frac{b_{12}}{8} [-3{\sigma_z}^{(1)} + {\sigma_z}^{(2)} + {\sigma_z}^{(3)} + {\sigma_z}^{(4)} ] \nonumber \\
	+ & \frac{b_{23}}{4} [-{\sigma_z}^{(1)} - {\sigma_z}^{(2)} + {\sigma_z}^{(3)} + {\sigma_z}^{(4)} ] \nonumber \\ 
	+ &\frac{b_{34}}{8}  [-{\sigma_z}^{(1)} - {\sigma_z}^{(2)} - {\sigma_z}^{(3)} +3 {\sigma_z}^{(4)} ] \nonumber \\	
    + \frac{1}{4}\sum_{\langle i j \rangle}&J_{ij}(\bm{\epsilon})\bm{\sigma} ^{(i)} \cdot \bm{\sigma} ^{(j)},
\end{align}
where the term involving $B_\mathrm{G}$ can be ignored for the qubit dynamics (since $\sum_{i=1}^4 {\sigma_z}^{(i)}$ is zero on the qubit subspace).

\section{Data vs. readout noise: anisotropic Random Bond Ising Model (aRBIM)}
\label{sec:aRBIM} 

As established in \cite{Dennis2001, Wang2003}, the maximum-likelihood decoding problem for the toric code subjected to phenomenological noise can be mapped onto the evaluation of the partition function of a classical spin model. The spin model is disordered, depending on quenched randomness which is due to the random outcomes of syndrome measurements. As mentioned in the main text, the decoding of the repetition code with noisy parity checks with phenomenological error rate $q=p$ is isomorphic to the surface code with data qubit error rate $p$ and perfect readout. Both models map onto the 2D Random Bond Ising model (RBIM), which has the Hamiltonian
\begin{equation}
H = -\sum_{\langle i j \rangle} J_{ij} S_i S_j
\end{equation}
with classical spin variables $S_i=\pm 1$ on a square 2D lattice. The sign of the coupling is drawn from a bimodal probability distribution
\begin{align}
  J_{ij} = +J \quad \text{ with } 1-p, \quad   J_{ij} = -J \quad \text{ with } p.
\end{align}
This entails antiferromagnetic bond defects with probability $p$ ($p$ is usually called the bond concentration). Antiferromagnetic bonds form strings, whose endpoints are called Ising vortices. These endpoints correspond to the syndrome defects as defined earlier and AFM bond strings correspond to possible error strings. 

The decoding problem can then be phrased as finding recovery strings with the same endpoints as the error string and asking whether they fall in the same homology class or not. Let us imagine some error string the qubits have suffered. This string is in actuality hidden to us, the syndrome only reveals the endpoints of strings. We (the decoder) construct ``candidate'' recovery strings to explain the syndrome that is seen and recover from the error the qubits have suffered. The combination of error string plus recovery string necessarily forms a loop, the question of logical correctness is whether and with what likelihood this loop is homologically non-trivial. 

This argument is then turned around, namely if the probability of being in the same homology class goes to unity in the thermodynamic limit, the likelihood of making a logical error goes to zero. To make the connection to the classical spin model, the recovery strings can be viewed as being created by thermal excitations. As long as error string plus recovery string form a loop with trivial homology, they are the boundary of a region, which is a domain wall. The phase diagram of the spin model has two axes, the bond concentration $p$ and the temperature $T = \frac{1}{\beta}$. Since the error string (setting the bond defects) and the recovery string are actually drawn from the same distribution, the Boltzmann weight and the bond concentration have to be identified as
\begin{equation}
  e^{2\beta J} = \frac{p}{1-p},
\end{equation}
which is called the Nishimori line in the two-dimensional $\beta-p$ plane . The correctability condition is thus, --in the statistical mechanics picture--, translated into the condition that domain walls remain localized, which is the case in the ferromagnetic phase of the Random Bond Ising model. The threshold of the code is given by the point where the phase boundary of the ferromagnetic to paramagnetic transition crosses the Nishimori line (at $p_c \approx 0.11$ \cite{Wang2003}).

\subsection{Anisotropic RBIM: Takeda-Nishimori conjecture}
We review the lesser known case $p \neq q$, corresponding to the fact that time-like AFM bonds are created with a different probability than space-like bonds. This maximum likelihood decoding problem can then be mapped to the evaluation of the partition function of an anisotropic Random Bond Ising model
\begin{equation}
H = -\sum_{\langle i j \rangle \in \text{vertical}} J_{ij}^v S_i S_j -\sum_{\langle i j \rangle \in \text{horizontal}} J_{ij}^h S_i S_j
\end{equation}
with
\begin{align}
  J_{ij}^v = J^v \quad \text{ with } 1-p, \quad  \quad J_{ij}^h = J^h \quad \text{ with } 1-q\\
  J_{ij}^v = -J^v \quad \text{ with } p, \quad \quad J_{ij}^h = -J^h \quad \text{ with } q.
\end{align}
and the Nishimori conditions
\begin{equation}
  e^{2\beta J^h} = \frac{p}{1-p},\quad e^{2\beta J^v} = \frac{q}{1-q}.
\end{equation}
This model has been analyzed by Takeda et al.~\cite{Takeda2005}. The spin model has four free parameters, the two bond concentrations $p$ and $q$ and the two coupling strengths ($J^h$ and $J^v$). Enforcing the two Nishimori conditions projects the four dimensional configuration space down onto a two-dimensional plane with the bond concentrations $p$ and $q$ as free parameters. This plane is called the Nishimori sheet (analogous to the well-known Nishimori line), on which the criticality condition, being a relation between $p$ and $q$, corresponds to a line, the critical line marking the phase boundary between error suppressing (ferromagnetic) and error enhancing (paramagnetic) bond concentrations, see Fig.~\ref{fig:memory-phase} in the main text.
\bibliography{thebibliography}{}
\end{document}